\documentclass[a4paper,11pt]{article}
\usepackage{jcappub} 

\usepackage{orcidlink}
\usepackage{graphicx}	
\usepackage{amsmath}	
\usepackage{booktabs}
\usepackage{multirow}
\usepackage{xcolor}
\usepackage{afterpage}
\usepackage[notquote]{hanging}
\usepackage{array}
\usepackage{arydshln}
\usepackage{adjustbox}
\usepackage{rotating,tabularx}
\usepackage{xspace}
\usepackage{multirow}

\usepackage{aas_macros}

\newcommand{\hinvmpc}{h^{-1}\,{\rm Mpc}}

\newcommand{\lcdm}{$\Lambda$CDM}

\title{\boldmath Fiducial-Cosmology-dependent systematics for the DESI 2024 Full-Shape Analysis}

\emailAdd{rafaela.gsponer@epfl.ch}
\affiliation{Affiliations are in Appendix \ref{sec:affiliations}}

\author[1]{{R.~Gsponer}\orcidlink{0000-0002-7540-7601},}
\author[2,3]{{S.~Ramirez-Solano}\orcidlink{0009-0006-2096-7974},}
\author[2,3]{{F.~Rodr{\'\i}guez-Mart{\'\i}nez},}
\author[2]{{M.~Vargas-Maga\~na}\orcidlink{0000-0003-3841-1836},}
\author[4]{{S.~Novell-Masot},}
\author[5]{{N.~Findlay}\orcidlink{0009-0007-0716-3477},}
\author[6,7,4]{{H.~Gil-Mar\'in}\orcidlink{0000-0003-0265-6217},}
\author[8]{{P.~Zarrouk}\orcidlink{0000-0002-7305-9578},}
\author[5]{{S.~Nadathur}\orcidlink{0000-0001-9070-3102},}
\author[1,9]{{A.~Rocher}\orcidlink{0000-0003-4349-6424},}
\author[10]{{S.~Brieden}\orcidlink{0000-0003-3896-9215},}
\author[11]{{A.~P\'erez-Fern\'andez}\orcidlink{0009-0006-1331-4035},}
\author[12]{{J.~Aguilar},}
\author[13]{{S.~Ahlen}\orcidlink{0000-0001-6098-7247},}
\author[14,15]{{D.~Bianchi}\orcidlink{0000-0001-9712-0006},}
\author[16]{{D.~Brooks},}
\author[7,17]{{F.~J.~Castander}\orcidlink{0000-0001-7316-4573},}
\author[12]{{T.~Claybaugh},}
\author[12]{{A.~Cuceu}\orcidlink{0000-0002-2169-0595},}
\author[2]{{A.~de la Macorra}\orcidlink{0000-0002-1769-1640},}
\author[9]{{A.~de~Mattia}\orcidlink{0000-0003-0920-2947},}
\author[18]{{Arjun~Dey}\orcidlink{0000-0002-4928-4003},}
\author[16]{{P.~Doel},}
\author[19]{{A.~Font-Ribera}\orcidlink{0000-0002-3033-7312},}
\author[20,21]{{J.~E.~Forero-Romero}\orcidlink{0000-0002-2890-3725},}
\author[7,5,17]{{E.~Gaztañaga}\orcidlink{0000-0001-9632-0815},}
\author[12,22]{{S.~Gontcho A Gontcho}\orcidlink{0000-0003-3142-233X},}
\author[23]{{G.~Gutierrez},}
\author[12]{{J.~Guy}\orcidlink{0000-0001-9822-6793},}
\author[24]{{C.~Hahn}\orcidlink{0000-0003-1197-0902},}
\author[25,9]{{H.~K.~Herrera-Alcantar}\orcidlink{0000-0002-9136-9609},}
\author[26,27,28]{{K.~Honscheid}\orcidlink{0000-0002-6550-2023},}
\author[29]{{C.~Howlett}\orcidlink{0000-0002-1081-9410},}
\author[30,31]{{D.~Huterer}\orcidlink{0000-0001-6558-0112},}
\author[32]{{M.~Ishak}\orcidlink{0000-0002-6024-466X},}
\author[18]{{R.~Joyce}\orcidlink{0000-0003-0201-5241},}
\author[33]{{R.~Kehoe},}
\author[34]{{D.~Kirkby}\orcidlink{0000-0002-8828-5463},}
\author[12]{{T.~Kisner}\orcidlink{0000-0003-3510-7134},}
\author[12]{{A.~Kremin}\orcidlink{0000-0001-6356-7424},}
\author[16]{{O.~Lahav},}
\author[35]{{C.~Lamman}\orcidlink{0000-0002-6731-9329},}
\author[12]{{M.~Landriau}\orcidlink{0000-0003-1838-8528},}
\author[8]{{L.~Le~Guillou}\orcidlink{0000-0001-7178-8868},}
\author[12]{{M.~E.~Levi}\orcidlink{0000-0003-1887-1018},}
\author[9]{{C.~Magneville},}
\author[36,19]{{M.~Manera}\orcidlink{0000-0003-4962-8934},}
\author[18]{{A.~Meisner}\orcidlink{0000-0002-1125-7384},}
\author[37,19]{{R.~Miquel},}
\author[38]{{J.~Moustakas}\orcidlink{0000-0002-2733-4559},}
\author[39]{{E.~Mueller},}
\author[9,12]{{N.~Palanque-Delabrouille}\orcidlink{0000-0003-3188-784X},}
\author[40,41,42]{{W.~J.~Percival}\orcidlink{0000-0002-0644-5727},}
\author[43]{{F.~Prada}\orcidlink{0000-0001-7145-8674},}
\author[44]{{I.~P\'erez-R\`afols}\orcidlink{0000-0001-6979-0125},}
\author[45]{{G.~Rossi},}
\author[46,47,48]{{L.~Samushia}\orcidlink{0000-0002-1609-5687},}
\author[49]{{E.~Sanchez}\orcidlink{0000-0002-9646-8198},}
\author[12]{{D.~Schlegel},}
\author[30,31]{{M.~Schubnell},}
\author[50]{{H.~Seo}\orcidlink{0000-0002-6588-3508},}
\author[12]{{J.~Silber}\orcidlink{0000-0002-3461-0320},}
\author[18]{{D.~Sprayberry},}
\author[31]{{G.~Tarl\'{e}}\orcidlink{0000-0003-1704-0781},}
\author[18]{{B.~A.~Weaver},}
\author[51]{{C.~Zhao}\orcidlink{0000-0002-1991-7295},}
\author[12]{{R.~Zhou}\orcidlink{0000-0001-5381-4372},}
\author[52]{{H.~Zou}\orcidlink{0000-0002-6684-3997},}

\abstract{We assess the impact of the fiducial cosmology choice on cosmological inference from full-shape (FS) fits of the galaxy power spectrum in the DESI 2024 Data Release 1 (DR1). Using a suite of AbacusSummit DR1 mock catalogues based on the Planck 2018 best-fit cosmology, we quantify potential systematic shifts introduced by analysing the data under five secondary cosmologies— featuring variations in matter density, thawing dark energy, higher effective number of neutrino species, reduced clustering amplitude, and the DESI DR1 BAO best-fit $w_0w_a$CDM cosmology —relative to DESI’s baseline Planck 2018 cosmology.  We investigate two complementary FS analysis approaches: full-modelling (FM) and ShapeFit (SF), each with distinct sensitivities to the assumed fiducial model. Across all tracers, we find for FM that systematic shifts induced by fiducial cosmology mismatches remain well below the DESI DR1 statistical uncertainties, with maximum deviations of 0.22$\sigma_\mathrm{DR1}$ in \lcdm{} scenarios and 0.12$\sigma_\mathrm{DR1+SN}$ when including SN Ia mock data in extended $w_0w_a$CDM fits. For SF, the shifts in the compressed parameters remain below $0.45\sigma_\mathrm{DR1}$ for all tracers and cosmologies. } 

\begin{document}
\vspace*{-2cm}
\maketitle
\flushbottom

\section{Introduction}
\label{sec:Introduction}
Over the past decades, galaxy redshift surveys have played a fundamental role in mapping the large-scale distribution of matter in the Universe. Pioneering spectroscopic programs, including the CfA Redshift Survey~\citep{CfA1983} and the 2dF Galaxy Redshift Survey (2dFGRS,~\citep{2dF2001}), sought to collect the angular positions and redshifts of galaxies, thereby constructing the first three-dimensional clustering maps. The clustering pattern of galaxies is sensitive to cosmic expansion, particularly through the baryon acoustic oscillation (BAO) feature, which results from sound waves propagating in the early Universe. These sound waves left a characteristic imprint at the sound horizon scale, $r_d\sim150~\rm Mpc$, in the distribution of matter at later times. Additionally, the growth of structure can be probed by studying the impact of peculiar velocities on galaxy clustering along the line of sight. Constraints on these two features can be obtained through individual tests, such as the Alcock-Paczynski (AP) test~\citep{Alcock1979} and Redshift-Space Distortions (RSD)~\citep{Kaiser1987,Hamilton:1992zz}. The AP test exploits the fact that assuming an incorrect background cosmology distorts radial and angular scales differently when converting redshifts into distances. By analysing isotropic features such as the BAO signal, the background cosmology can be constrained by measuring scaling parameters across and along the line of sight: $\alpha_\parallel, \alpha_\perp$. These parameters are pivotal in BAO analyses due to their direct interpretations in terms of the Hubble distance $D_{H}(z)$ and comoving angular diameter distance $D_M (z)$, respectively. Similarly, the RSD effect manifests as a clustering enhancement along the line of sight due to peculiar motions of galaxies. This effect is quantified through the logarithmic growth rate $f$ and the amplitude of matter fluctuation $\sigma_8$, often combined as $f \sigma_8$. RSD measurements offer a powerful test of gravity on cosmological scales, helping to distinguish between models of cosmic structure formation and potential modifications to General Relativity. 

CfA and 2dFGRS enabled the testing of fundamental cosmological principles, such as the detection of BAO~\citep{Percival2001, Cole2005} as a probe of the cosmic expansion history and constraints on gravity through RSD~\citep{Peacock2001, Guzzo2008}. Advances in imaging techniques and computational processing allowed the Sloan Digital Sky Survey (SDSS,~\citep{SDSS2006}) to improve on these initial spectroscopic surveys, ultimately measuring more than 2 million galaxies over the span of nearly two decades. The initial detection of BAO using SDSS data was accomplished by~\citep{Eisenstein2005}.  Subsequent projects, such as the Baryon Oscillation Spectroscopic Survey (BOSS)~\citep{BOSS2013} and its extension (eBOSS)~\citep{eBOSS2016}, firmly established BAO as a standard ruler~\citep{Eisenstein2005} to obtain geometrical information of the Universe and RSD as a dynamical test of dark energy.

The Dark Energy Spectroscopic Instrument (DESI)~\citep{DESIWhitepaper2013, DESI2016.InstrumentationMethods, DESI2022.Overview,DESICollaboration2016a,FocalPlane.Silber.2023,Corrector.Miller.2023,Spectro.Pipeline.Guy.2023,FiberSystem.Poppett.2024,SurveyOps.Schlafly.2023,LRG.TS.Zhou.2023,DESI2023a.KP1.SV,DESI2023b.KP1.EDR,DESI2024.II.KP3, DESI_DR1} represents the next step in this field. Designed to explore the expansion history and the nature of dark energy, DESI is a stage-IV spectroscopic survey targeting nearly 40 million galaxies and quasars across $0 < z< 4$. Over a span of five years, DESI will cover approximately $14,000$~deg$^2$ of the sky. A key innovation of DESI is the incorporation of 5,000 robotic fibre positioners, a feature that enables the observation of multiple targets simultaneously and significantly enhances the efficiency of the survey. In its first year alone (Data Release 1; DR1), DESI measured over 4.7 million unique galaxies and quasars redshifts, thereby achieving the most precise BAO~\citep{DESI2024.III.KP4, DESI2024.IV.KP6, DESI2024.VI.KP7A} and RSD~\citep{DESI2024.V.KP5,DESI2024.VII.KP7B} measurements to date. In addition to fitting individual features of the power spectrum, as it is done e.g. in the BAO analysis of DESI~\citep{DESI2024.III.KP4}, the full-shape analysis of DESI~\citep{DESI2024.V.KP5} takes a complementary approach by modelling a broad range of scales of the power spectrum. This approach provides valuable insights into the growth of structure as a function of redshift, which is highly sensitive to the nature of dark energy, modified gravity, and the total matter content of the Universe. Moreover, it enables the comparison of alternative gravity models that predict the same expansion history but differ in their description of structure formation. Cosmological constraints from the full-shape analysis can be obtained in two ways: by directly constraining the parameters of a specific cosmological model (Full-Modelling), or by first constraining a reduced set of parameters (compressed) that capture essential information about the model and subsequently interpreting these in terms of cosmological parameters of a chosen model.

This paper is part of a broader series of studies that investigate various potential sources of systematic uncertainty in the DESI DR1 full-shape analysis~\citep{DESI2024.V.KP5}. Each study focuses on a different aspect of the analysis, including theoretical and modelling systematics \citep{KP5s1-Maus,KP5s2-Maus,KP5s3-Noriega,KP5s4-Lai,KP5s5-Ramirez}, Halo Occupation Distribution (HOD) systematics \citep{KP5s7-Findlay}, imaging systematics \citep{KP5s6-Zhao}, spectroscopic systematics \citep{KP3s4-Yu,KP3s3-Krolewski}, fiber collision effects \citep{KP3s5-Pinon}, and covariance matrices \citep{KP4s6-Forero-Sanchez,KP4s8-Alves,KP4s7-Rashkovetskyi}. This particular study examines the systematics associated with the choice of a fiducial set of cosmological parameters assumed in the full-shape analysis. 
This fiducial cosmology choice affects the full-shape analysis in two main ways. First, galaxy positions are transformed from observed coordinates (redshift and angular positions) to comoving coordinates using an assumed redshift-to-distance relation. Here, the set of cosmological parameters that sets the comoving distance as a function of redshift, denoted as $D_\mathrm{grid}(z)$, is referred to as the \textit{grid} cosmology. If the fiducial model deviates from the true cosmology, anisotropic distortions are introduced to the power spectrum, shifting the BAO peak and altering the broadband shape of the power spectrum. Second, in compressed full-shape analyses, a fixed fiducial cosmology is used to generate a power spectrum \textit{template} for comparison with the data. Any discrepancy between the fiducial template and the actual cosmology can result in biases in the inferred parameters. For earlier studies on the impact of the \textit{template} cosmology in compressed full-shape analysis within the context of spectroscopic surveys, we direct the reader to~\citep{GilMarinBOSS:2016,NeveuxeBOSS2020, GilMarineBOSS2020,deMattiaeBOSS2020,Brieden_2021JCAP}, and for BAO analyses to~\cite{VargasMagana2018,Carter2020}.  This paper aims at quantifying the effects of the chosen \textit{grid} and \textit{template} cosmologies, assessing any potential systematic shifts introduced by fiducial cosmology assumptions in the DESI full-shape clustering analysis, and providing crucial insights for the interpretation of DESI’s cosmological constraints.

The paper is structured as follows. Section~\ref{sec:Model} provides an overview of the theoretical modelling used in the DESI full-shape analysis. Section~\ref{sec:FiducialCosmology} introduces the fiducial cosmologies tested in this paper. In Section~\ref{sec:Data}, we present the mock catalogues used for our tests. Section~\ref{sec:Method} details the methodology and Section~\ref{sec:Results} presents the results, quantifying the systematic shifts introduced by different fiducial cosmology choices. Section~\ref{sec:Conclusion} provides conclusions and discusses implications for future years of the DESI full-shape analysis.

\section{Modelling of the full-shape Power Spectrum}
\label{sec:Model}
In recent years, significant effort has gone into extending the theoretical modelling of the matter power spectrum into the mildly non-linear regime, thereby enabling access to a wider range of modes. These advancements are founded on the pillars of Standard Perturbation Theory (SPT), which has been expanded into an effective field theory approach to address challenges faced by conventional methods. The theoretical framework, which encompasses the mildly non-linear regime where small-scale effects can still be treated perturbatively, is referred to as the Effective Field Theory of Large Scale Structure (EFTofLSS)~\citep{Baumann:2010tm, Carrasco:2012cv}. EFTofLSS treats dark matter as an imperfect fluid, where small-scale non-linearities introduce dissipative effects and non-negligible anisotropic stress into the evolution of long-wavelength perturbations. This is achieved by coarse-graining the fluid equation at a smoothing scale $\Lambda$. The impact of the unknown small-scale physics beyond the cutoff scale $\Lambda$ on larger scales is captured in an effective stress-energy tensor $\sigma_{ij}$ which introduces counterterm parameters into the theory. At the level of the 1-loop redshift-space power spectrum, the EFTofLSS predictions for the dark matter field are mapped to the galaxy field through three distinct categories of nuisance parameters $\theta_\mathrm{nuis.}$: the galaxy bias parameters, which describe the relationship between the dark matter field and the biased tracer field; counter terms, which encode the uncertainity from our inability to model small-scale physics; and stochastic parameters, which account for discrepancies between the observed galaxy field and its expected value.

The process of extracting information from the theoretical modelling of the power spectrum and comparing it to observational data can be approached in two ways. The ShapeFit approach involves using compressed parameters, which inherently constrain and define features within the clustering signal, such as the scaling parameters of the BAO peak, the growth of structure parameter, or the overall slope of the power spectrum. The other approach, which we refer to as `Full-Modelling', involves working with the parameters of a chosen cosmological model, where the most prominent example is the standard \lcdm{} model, characterised by parameters such as $\{h, \Omega_m, A_s \dots \}$. Below, we provide a brief overview of these two approaches.

\subsection{Full-Modelling}
\label{sec:fullmodelling}
In the Full-Modelling (FM) approach, a cosmological model with a fixed set of cosmological parameters $\theta_\mathrm{cosmo.} = \{h, \Omega_m, A_s \dots \}$ is assumed and the linear matter power spectrum is predicted by using an Einstein-Boltzmann code. This is then input into the one-loop redshift-space galaxy power spectrum prediction from EFTofLSS. The total number of free parameters depends on both the cosmological parameters $\theta_\mathrm{cosmo.}$ and the EFTofLSS nuisance parameters $\theta_\mathrm{nuis.}$. The full power spectrum $P(k, \mu)$ is then compared to the data, incorporating not just the BAO and RSD signals but also broadband information such as the slope of the power spectrum. In the context of a cosmological analysis, the non-linear power spectrum is calculated at each step in terms of the newly proposed $\theta_\mathrm{cosmo.}$ and $\theta_\mathrm{nuis.}$. 

In order to account for distortions resulting from the assumption of fiducial cosmology (the \textit{grid} cosmology) when converting catalogue redshift information to comoving distances, the same distortion is applied to the theoretical predictions for each newly proposed set of $\theta_\mathrm{cosmo}$. This results in the scaling of the true wavenumbers parallel and perpendicular to the line of sight by the scaling parameters:
\begin{align}
    q_\perp(z) &= \frac{D_M(z)}{D_M^\mathrm{grid}(z)}\ , \label{eq:AP_FM1}\\
    q_\parallel(z) &= \frac{D_H(z)}{D_H^\mathrm{grid}(z)}\ ,
    \label{eq:AP_FM2}
\end{align}
where $D_M(z)$ and $D_H(z) \equiv c/ H(z)$ describe the comoving angular-diameter distance and the Hubble distance as a function of the Hubble rate $H(z)$, respectively.  The super-script `grid' stands for the parameters evaluated in the fiducial \textit{grid }cosmology. Note that since comoving distances are typically expressed in terms of $\hinvmpc$, (where $H_0 \equiv 100~h$~km/s/Mpc) only the $E(z)=H(z)/H_0$ function affects the fiducial $D_M(z)$ and $D_H(z)$ distances set by the \textit{grid} cosmology.

\subsection{ShapeFit}
\label{sec:shapefit}
The compression approach aims at extracting the information from specific features of the power spectrum $P(k,\mu)$ into physically meaningful observables in a nearly lossless and model-independent way. A similar philosophy underlies template-based BAO analyses, where the BAO signal is encapsulated in the scaling parameters $\alpha_\parallel$ and $\alpha_\perp$. 
Since the compressed approach prioritises model-independence for these parameters, necessitating the introduction of a reference scale. In standard BAO analyses, this reference scale is the sound horizon at the drag epoch, $r_\mathrm{d}$,  which sets the characteristic scale of the linear power spectrum template and enables a clear distinction between early-time physics and late-time geometric information.
As a result, the scaling parameters are expressed as the ratio of the true distances $D_M, D_H$ over their fiducial values $D_M^\mathrm{grid},D_H^\mathrm{grid}$  measured in units of the standard ruler $r_\mathrm{d}$ for a fixed \textit{template} cosmology:
\begin{align}
    \alpha_\perp(z) &= \frac{D_M(z)r_\mathrm{d}^\mathrm{temp.}}{D_M^\mathrm{grid}(z) r_\mathrm{d}}\ , \\
    \alpha_\parallel(z) &= \frac{D_H(z) r_\mathrm{d}^\mathrm{temp.}}{D_H^\mathrm{grid}(z) r_\mathrm{d}}\ .
    \label{eq:AP_SF}
\end{align}
Strictly speaking, the geometrical distortions in the distance ratios affect the full power spectrum shape, whereas the $r_\mathrm{d}/r_\mathrm{d}^\mathrm{temp.}$ rescaling affects primarily the position of the BAO wiggles. In the ShapeFit analysis, the scaling parameters are used to model the dilation of the full power spectrum shape over a broader range of scales ($0.02<k\,[{\rm Mpc}^{-1}h ]<0.20$). This has been demonstrated to be a good approximation because the prominent BAO signal is the primary feature responsible for setting a distinct {\it reference} scale in the analysis \cite{Brieden_2023JCAP}.
In previous RSD analyses \citep{GilMarinBOSS:2016,BeutlerBOSS2017,GriebBOSS2017,SatpathyBOSS2017,SanchezBOSS2017, GilMarineBOSS2020, NeveuxeBOSS2020, deMattiaeBOSS2020,HoueBOSS2020,BautistaeBOSS2020}, additional broadband information was incorporated through the amplitude parameter $f \sigma_8$ on top of the scaling parameters. All these observables depend solely on late-time geometry and kinematics, with early-time information entering only through the drag horizon $r_\mathrm{d}$. Consequently, further early-time information, such as the primordial tilt $n_s$ or the impact of the transfer function on broadband features, is typically neglected.
The ShapeFit (SF) method~\citep{Brieden_2021JCAP,Brieden_2023JCAP} addresses this limitation by extending the traditional compressed RSD fit to include key early-time information from the transfer function through two effective parameters, $m$ and $n$. The shape parameter $m$ corresponds to the maximum slope at a chosen pivot scale $k_p = \pi /r_\mathrm{d}^{\rm temp.}$, encapsulating information about the epoch of matter-radiation equality via the transfer function. The slope parameter $n$ is scale-independent and directly interpretable in terms of the primordial scalar tilt, defined as $n = n_s - n_s^\mathrm{temp.}$, in the standard $\Lambda$CDM model. With these new parameters, the power spectrum can be expressed in terms of a fiducial power spectrum template and a slope rescaling:
\begin{align}
    \tilde{P}_\mathrm{lin}(k) = P^\mathrm{temp.}_\mathrm{lin}(k) \exp \left\{ \frac{m}{a} \tanh \left[ a \ln \left( \frac{k}{k_p}\right)\right] + n \ln \left( \frac{k}{k_p}\right) \right\},
    \label{eq:SFtemplate}
\end{align}
where $a$ governs the transition rate between the large-scale and small-scale limits. As calibrated in~\citep{Brieden_2021JCAP}, we fix this parameter to $a= 0.6$ in our analysis. 

In practice, the linear power spectrum, $\tilde{P}_{\rm lin}$, is computed using an Einstein-Boltzmann solver for a chosen and fixed \textit{template} cosmology, $P_{\rm lin}^{\rm temp.}$. It is then extended to the non-linear galaxy power spectrum in redshift space using the EFTofLSS framework. As demonstrated in~\citep{Catelan_1995MNRAS}, the perturbation kernels exhibit only a weak dependence on the assumed cosmology,
allowing to compute the one-loop contributions only once at the reference cosmology
following the procedure described in Appendix B of \cite{KP5s3-Noriega}, \cite{KP5s5-Ramirez} and Appendix D of \cite{Brieden_2021JCAP}. 
This approximation is not adopted in our implementation (\texttt{desilike}), which instead performs all relevant integrals explicitly.
The template is subsequently modified according to the newly introduced SF parameters $m$ and $n$ before it is rescaled by the classic scaling parameters $\alpha_\perp, \alpha_\parallel$.  In the SF framework, the conventional amplitude parameter $\sigma_8$ is replaced by the parameter $\sigma_{s8}$. This distinction is made to highlight that $\sigma_{8}$ now depends on $m$ and to reflect the fact that the definition of the $8~\hinvmpc$ scale becomes ambiguous when $h$ is not fixed by a known cosmology. For a detailed definition of $\sigma_{s8}$, see~\citep{Brieden_2021JCAP,Brieden_2023JCAP}. The re-definition of $\sigma_{s8}$ only affects the interpretation of the SF results, while the fitting process (and modelling used therein) remains the same. In particular, $\sigma_{s8}$ will be equal to $\sigma_8$ if the true value of $r_\textrm{d}$ coincides with the fiducial choice, $r_\textrm{d}^\textrm{temp.}$, and the best-fit values for $\alpha_\parallel$ and $\alpha_\bot$ are 1. 
The prediction of the galaxy power spectrum, based on the compressed parameters $\theta_\mathrm{compr.} = \{\alpha_\parallel, \alpha_\perp, f\sigma_{s8}, m, n \}$ along with the nuisance parameters $\theta_\mathrm{nuis.}$ of the EFTofLSS theory, can then be compared to the observed data. In a final step, the constraints on the SF parameters  $\theta_\mathrm{compr.}$ can be interpreted in terms of the cosmological parameters of a specific model by comparing the log-likelihood values for the theoretical predictions of $\theta_\mathrm{compr.}$ according to a chosen cosmological model and its parameter values at each step of the cosmological analysis.

\section{Choice of the Fiducial Cosmology}

\label{sec:FiducialCosmology}
In this section, we investigate the impact of fiducial cosmology on the analysis within both the full-modelling and compressed approaches. The role of the fiducial cosmology varies depending on the approach: in the full-modelling case, it enters through the \textit{grid} cosmology, whereas in the compressed approach, it affects both the \textit{grid} and \textit{template} cosmologies. We first describe how the fiducial cosmology assumption is incorporated in each approach, before giving an overview of the different fiducial cosmology choices explored in this paper. These cosmologies are selected to evaluate how deviations between the true underlying cosmology of the Universe and a chosen fiducial cosmology affect the analysis of large-scale structure data.\\

\textbf{Grid cosmology}: To impose a coordinate system, a fiducial cosmology is chosen when galaxy positions measured by DESI are converted from redshift and angles into Cartesian coordinates. This transformation is based on the distance-redshift relation, where the assumption of a fiducial cosmology enters through the functional form of the Hubble rate in the definition of the comoving distance $D^\mathrm{grid}(z)$. Within \lcdm{}, the comoving distance is primarily influenced by today's expansion rate $H_0 = 100~h$~km/s/Mpc, as well as by the matter density at present $\Omega_{\rm m}$. By expressing the distances in units of $\hinvmpc$  we reduce this dependence on only the unnormalised expansion history $E(z)$, and hence to $\Omega_m$ for \lcdm{}. Choosing an incorrect \textit{grid} cosmology affects both the full-modelling and ShapeFit analyses, manifesting as alterations in the scaling parameters, which will no longer be equal to unity. Moreover, if $q_\parallel \neq q_\perp$, this mismatch introduces the AP effect, generating a spurious anisotropic signal.\\

\textbf{Template cosmology:} In the compressed full-shape analysis, the choice of fiducial cosmology defines the setup of the power spectrum template. The initial fiducial linear power spectrum, $P_{\rm lin}^{\rm temp.}(k) \equiv P_{\rm lin}(k, \theta^{\rm temp.})$, is set by baseline cosmological parameters such as the primordial tilt $n_s$. The sound horizon at the drag epoch, $r_\mathrm{d}$, is a derived parameter from this cosmology and governs the location of the BAO features. Additionally, the shape transformation applied to generate the rescaled template $\tilde{P}_{\rm lin}(k)$ involves the slope parameter $m$, as described in Eq.~\eqref{eq:SFtemplate}.\\

\begin{figure}[htb]
    \centering
    \includegraphics[width=1.\linewidth]{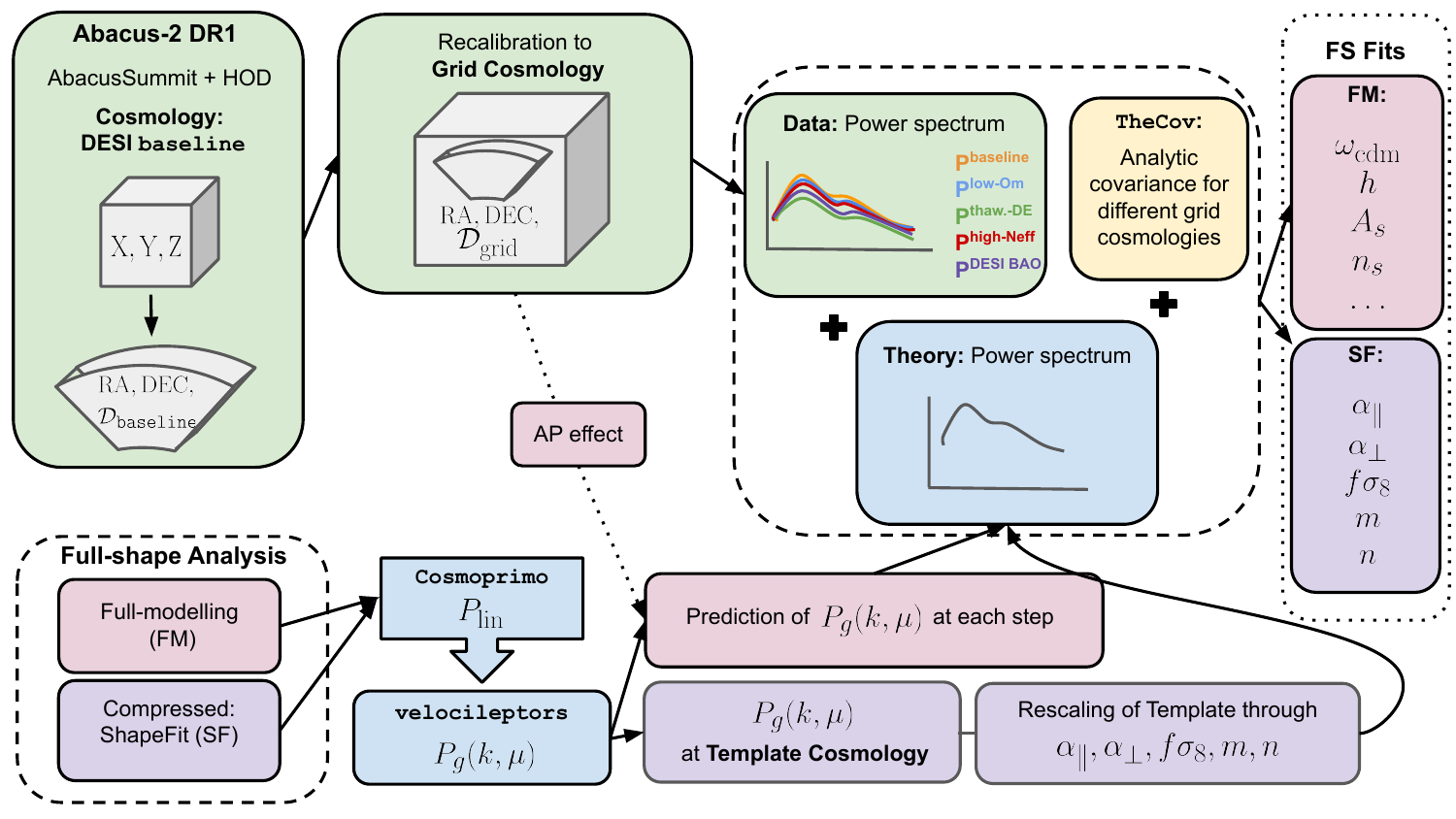}
    \caption{Schematic flowchart illustrating the DESI Full-Shape analysis and the role of the fiducial cosmology assumption. The green boxes represent steps related to the catalogue creation, including distance calibration and summary statistics computation. The blue boxes indicate the generation of the theoretical non-linear galaxy power spectrum for a given set of trial cosmological parameters. This trial set of cosmological parameters is represented in pink boxes for the FM approach, and in purple boxes for the compressed SF approach. The yellow box marks the step where analytic covariance is computed and incorporated. Finally, the red and orange boxes represent the best-fitting cosmological parameters from the FM and SF analyses, respectively, as obtained by fitting the data.
    }
    \label{fig:flowchart}
\end{figure}

Figure~\ref{fig:flowchart} illustrates where the assumption of the fiducial cosmology enters the analysis for the FM and SF approaches. While it is typical to choose the same fiducial cosmology for both the \textit{grid} and \textit{template} in compressed analyses, it is, in principle, possible to use different cosmologies for each. For simplicity, we assume that the \textit{grid} and \textit{template} cosmologies are identical throughout this paper, unless otherwise stated.  A detailed discussion of the impact of varying only the \textit{grid} or \textit{template} cosmology in the context of SF is provided in Section~\ref{sec:SFresults}. 

In this study, we investigate the impact of different fiducial cosmology choices by analysing galaxy mocks with a fixed true underlying cosmology. Specifically, we consider four cosmologies from the AbacusSummit suite~\citep{AbacusSummit} and the DESI DR1 BAO best-fit $w_0w_a$CDM cosmology~\citep{DESI2024.VI.KP7A}, and using the DESI \texttt{baseline} cosmology~\citep{DESI2024.III.KP4, DESI2024.V.KP5} as the true underlying model of the mocks. We then compare results obtained when adopting five alternative fiducial cosmologies against those using the correct \texttt{baseline} cosmology, assessing the systematic shifts introduced by an incorrect fiducial cosmology assumption. The \texttt{baseline} cosmology is identical to the primary AbacusSummit cosmology and corresponds to the mean estimate of the \textit{Planck}~2018 TT,TE,EE$+$lowE$+$lensing \lcdm{} chains~\citep{Planck2018}. While all cosmologies share a fixed reionisation optical depth of $\tau = 0.0544$ and include one massive neutrino species, the secondary cosmologies differ in key parameters, such as the cold dark matter density $\omega_\mathrm{cdm}$, the number of ultra-relativistic species $N_\mathrm{ur}$, the dark energy equation of state $(w_0, w_a)$, and the amplitude  of matter fluctuations $\sigma_8$. The \texttt{baseline} cosmology assumes $N_\mathrm{ur} = 2.0328 $ and includes one massive neutrino species with $\omega_\nu = 0.00064420$.  The \texttt{low-$\Omega_\mathrm{m}$}  features a lower matter density, with $\omega_\mathrm{cdm} = 0.1134$ and $\Omega_m = 0.2761$, and corresponds to the mean of the WMAP9+ACT+SPT \lcdm{} chains~\citep{Calabrese_2017}. The $\texttt{thawing-DE}$ cosmology includes a dynamical dark energy model $w_0w_a$CDM, with $w_0 = -0.7$ and $w_a = -0.5$~\citep{Chevallier_2001,Linder_2003}. The \texttt{high-$N_\mathrm{eff}$} cosmology features a higher effective number of neutrino species, with $N_{\rm eff} = 3.70$, corresponding to an effective number of ultra-relativistic of $N_{\rm ur} = 2.6868$. The \texttt{low-$\sigma_8$} cosmology represents the \texttt{baseline}  cosmology, but with a reduced clustering amplitude $\sigma_8 = 0.75$, a $7.7\%$ decrease from the fiducial value~\citep{AbacusSummit} due to a lowered value of $A_s$.
While the \texttt{low-$\sigma_8$} cosmology directly influences the overall amplitude of the \textit{template}, it does not affect the \textit{grid} cosmology since $\sigma_8$ only affects the density fluctuations, not the cosmic expansion history used to calculate comoving distances. The \texttt{low-$\sigma_8$} cosmology  is therefore only considered in the SF analysis. 
Lastly, we consider as our final secondary fiducial cosmology the best-fit \texttt{DESI BAO} $w_0w_a$ model from the DESI DR1 analysis~\citep{DESI2024.VI.KP7A}. This best-fit cosmology is derived from a combination of datasets that includes DESI DR1 BAO measurements~\citep{DESI2024.III.KP4}, temperature and polarization data from Planck~\citep{Planck2018}, CMB lensing from the joint Planck+ACT analysis~\citep{Planck2022lensing,ACT2024_1,ACT2024_2,ACT2024_3}, and the Dark Energy Survey (DES) Year 5 Supernovae Ia sample~\citep{DESY5SN2024}. The maximum a posteriori (MAP) point of the chain corresponds to $w_0 = -0.73$ and $w_a = -1.01$.
A summary of the individual cosmologies is provided in Table~\ref{tab:fid_cosmo}\footnote{We note that in other papers, the AbacusSummit cosmologies discussed here are referred to as \texttt{c000-c004} in the official AbacusSummit nomenclature, with \texttt{baseline} corresponding to \texttt{c000}. While this naming convention is commonly used in other DESI papers, a more descriptive naming scheme was employed here, where the names directly reflect the specific characteristics of each cosmology, such as \texttt{low-$\Omega_m$, thawing-DE, high-$N_\text{eff}$} and \texttt{low-$\sigma_8$}. }. These cosmologies align with the fiducial cosmologies considered in the companion BAO fiducial cosmology systematic paper~\citep{KP4s9-Perez-Fernandez}.\\

\begin{table}[!ht]
    \centering
           \resizebox{\columnwidth}{!}{
    \begin{tabular}{|l|c|c|c|c|c|c|c|c|c|c|}
    \hline
        Name  &  $\omega_{\rm b}$ & $\omega_{\rm cdm}$ & $\Omega_m$ & $h$ & $10^{9}A_s$ & $n_s$ & $N_{\rm ur}$ & $w_0$ & $w_a$ & $\sigma_8$\\
       \hline\hline 
       \texttt{baseline}&  $0.02237$ & $0.1200$ & $0.31519$ & $0.6736$ & $2.0830$ & $0.9649$ & $2.0328$ & $-1$ & $0$ & $0.8080$\\
       \texttt{Low-$\Omega_m$} & $0.02237$ & $0.1134$ & $0.27613$ & $0.7030$ & $2.0376$ & $0.9638$ & $2.0328$ & $-1$ & $0$ &$0.7768$\\
       \texttt{Thawing-DE}&  $0.02237$ & $0.1200$ & $0.36286$ & $0.6278$ & $2.3140$ & $0.9649$ & $2.0328$ & $-0.7$ & $-0.5$& $0.8082$\\
       \texttt{High-$N_{\rm eff } $}  & $0.02260$ & $0.1291$ & $0.29717$ & $0.7160$ & $2.2438$ & $0.9876$ & $2.6868$ & $-1$ & $0$ &$0.8552$\\
       \texttt{Low-$\sigma_8$}  & $0.02237$ & $0.1200$ & $0.31519$ & $0.6736$ & $1.7949$ & $0.9649$ & $2.0328$ & $-1$ & $0$ & $0.7500$\\
       \texttt{DESI BAO}  & $0.02243$ & $0.1197$ & $0.31585$ & $0.6724$ & $2.101$ & $0.9674$ & $2.0328$ & $-0.73$ & $-1.01$ & $0.8141$\\
\hline       
    \end{tabular}
    }
    \caption{Cosmological models employed in this paper as choices of fiducial cosmologies. The AbacusSummit mocks, which serve as the validation dataset, are based on an underlying true cosmology corresponding to the \texttt{baseline} row. Therefore, we investigate the impact of {\it relative} changes in the fiducial parameters with respect to the \texttt{baseline} model. Note that the values of $\sigma_8$ quoted here include the contribution of neutrinos. }
    \label{tab:fid_cosmo}
\end{table}

\section{Mock Data Sets}
\label{sec:Data}

\subsection{AbacusSummit-2 DR1}
\label{sec:AbacusSummit}

We generate mock catalogues based on the DESI \texttt{baseline} cosmology using the AbacusSummit suite of high-accuracy, high-resolution N-body simulations~\citep{AbacusSummit}. These simulations are designed to support large-scale structure analyses in the era of Stage-IV surveys and are specifically tailored to meet the requirements of the DESI clustering analysis~\citep{DESICollaboration2016a}. Subsequently, the distances in the mock catalogues are re-calibrated according to the five different \textit{grid} cosmologies (\texttt{baseline, low-$\Omega_\mathrm{m}$, thawing-DE, high-$N_\mathrm{eff}$, DESI BAO}) described in Section~\ref{sec:FiducialCosmology} in order to produce five sets of 25 individual power spectrum realisations. 

For our analysis, we use the ‘CutSky’ Abacus-2 DR1 mock set, as detailed in~\citep{KP3s8-Zhao}. This dataset consists of 25 realisations of $(2~h^{-1} \rm{Gpc})^3$ simulation boxes, each containing $6912^3$ dark matter particles with a particle mass of $2 \times 10^9~h^{-1}M_{\odot}$. While the full AbacusSummit suite encompasses 97 distinct cosmologies across 150 simulations, we focus on the Planck 2018~\citep{Planck2018} best fit \lcdm{} cosmology, which serves as the $\texttt{baseline}$ cosmology. The Abacus-2 DR1 ‘CutSky’ mocks are designed to replicate the geometry and radial distribution of the DESI DR1, ensuring realistic clustering properties. They were constructed by identifying halos using the CompaSO halo finder~\citep{compaso} and subsequently refined following~\citep{2021Bose}  to remove over-deblended halos and correctly merge physically associated halos that had been temporarily separated. Galaxies were then populated into these halos using the AbacusHOD model~\citep{2022AbacusHOD} for dark time tracers, while for bright time tracers, a halo tabulation method was applied to fit the halo occupation distribution (HOD) across different absolute magnitude threshold samples~\citep{EDR_BGS_ABACUS}. The mocks are calibrated to the DESI Early Data Release (EDR) data, as detailed in~\citep{EDR_HOD_ELG2023, EDR_HOD_LRGQSO2023,EDR_BGS_ABACUS}. To accurately match the full DESI DR1 footprint, simulation boxes were replicated following the method outlined in~\citep{KP3s8-Zhao}. We use simulation snapshots at different redshifts corresponding to specific tracers: $z = 0.2$ for BGS ($0.1 < z < 0.4$), $z = 0.5$, $0.8$ for LRGs ($0.4 < z < 1.1$), $z =  1.325$ for ELGs ($1.1 < z < 1.6$), and $z = 1.4$ for QSOs ($0.8 < z < 2.1$).  The box coordinates are then mapped to a CutSky geometry, with angular sky positions and observed redshifts derived from radial comoving distances relative to a chosen observer position and the \texttt{baseline} cosmology. The mock catalogues maintain the same geometry, redshift distribution $n(z)$, and completeness properties as the DESI data, which are applied through the DESI Large Scale Structure (LSS) pipeline~\citep{KP3s15-Ross}. 

In this analysis, we use the ‘complete’ Abacus-2 mocks, where all potential galaxy assignments are treated as observed, completeness weights are set to unity, and no priority veto masks are applied. Starting from the baseline cosmology mocks, as described above, we recalibrate distances from the observer using each of the five \textit{grid} cosmologies (\texttt{baseline, low-$\Omega_\mathrm{m}$, thawing-DE, high-$N_\mathrm{eff}$, DESI BAO}). The power spectrum statistics are computed using \texttt{pypower}\footnote{\url{https://github.com/cosmodesi/pypower}}, based on the estimator from~\citep{Yamamoto2006,Bianchi2015,Hand_2017}. The density field is interpolated onto a $512^3$ mesh using a triangular-shaped cloud (TSC) prescription~\citep{Jing2005,Sefusatti2016}, within an enclosing box size $L = \{4, 7, 9, 10\}~h^{-1} \rm{Gpc}$ for BGS, LRGs, ELGs and QSO, respectively. The power spectrum multipoles are initially binned with a width of $\Delta k = 0.001~\hinvmpc$, later re-binned to $\Delta k = 0.005~\hinvmpc$. This results in $5\times25$ distinct data vectors corresponding to different \textit{grid} cosmologies. In the following sections, we analyse these data vectors to assess the impact of the chosen \textit{grid} cosmology on the inferred cosmological parameters. For each \textit{grid} cosmology, we also compute the survey window matrix following the prescription of~\citep{Beutler_2021} using \texttt{pypower}.

\subsection{Analytical Covariance Matrices}
\label{sec:covariance}
We compute analytic covariance matrices for the pre-reconstructed power spectrum multipoles of the ‘complete’ Abacus-2 mocks described above. This is done using the code \texttt{thecov}\footnote{\url{https://github.com/cosmodesi/thecov}}, which implements the perturbative approach developed in~\citep{Wadekar:2019rdu} to compute the Gaussian contribution to the covariance of the galaxy power spectrum multipoles. A key feature of this approach is its efficient incorporation of the survey window function, treating it as a set of fixed kernels that can be precomputed using fast Fourier transforms (FFTs). This enables a fast evaluation of the covariance matrix while maintaining flexibility in varying cosmological and bias parameters. The method has been tested within the DESI analysis framework in~\citep{KP4s8-Alves}. To assess the limitation of considering only the Gaussian contribution, it has been further compared against mock-based covariance estimates in~\citep{KP4s6-Forero-Sanchez}, which found that the analytic covariance slightly underestimates the variance observed in the mocks. Nevertheless, the Gaussian approximation remains sufficiently accurate for the purposes of our analysis, where we are primarily interested in the shift of the maximum a posterior (MAP) values in cosmological parameters due to different choices of the fiducial cosmology. The input power spectrum used in the covariance calculation is the average of the 25 mock realisations. A separate covariance matrix is computed for each \textit{grid} cosmology.\\

\subsection{DR1-like errors}
\label{sec:DR1error}
For determining the shifts in the MAP values, we fit the mean of 25 realisations for each of the five \textit{grid} cosmologies using the appropriate analytical covariance matrix. Unless stated otherwise,this covariance is rescaled by a factor of 25, corresponding to the total volume of these realisations ($\mathrm{V}_{25}$). Nevertheless, the primary objective of this paper is to evaluate the shifts anticipated for realistic DESI DR1 uncertainties. To this end, we introduce the notation of the statistical uncertainty in DR1 as $\sigma_\mathrm{DR1}$. 

As in the official DR1 analysis, we rely on covariance matrices constructed from 1000 effective Zel'dovich approximation mock realisations (EZmocks;~\citep{Chuang2015}) to define the DR1 statistical uncertainty. These large ($6~h^{-1} \rm{Gpc})^3$ mocks enable the reproduction of the DR1 survey geometry without requiring replication of the periodic box. While the ‘complete’ Abacus-2 mocks assume no fiber collisions, the EZmocks incorporate fiber assignment using the `fast-fibreassign' method (FFA;~\citep{DESI2024.II.KP3,KP3s11-Sikandar}). This method also accounts for the effect of the $\theta$-cut, which mitigates fiber assignment incompleteness by discarding pair counts with angular separations smaller than $\theta = 0.05^{\circ}$~\citep{KP3s11-Sikandar}. Section 10.2 of~\citep{DESI2024.II.KP3} highlights that covariance matrices constructed from the EZmocks underestimate the variance observed in real DR1 data due to limitations in the FFA approximation. To account for this discrepancy, a rescaling factor is applied to the EZmock covariance for each tracer, calibrated based on the mismatch with the configuration-space DR1 covariance~\citep{KP4s7-Rashkovetskyi} (see table~7 in~\citep{DESI2024.V.KP5}). The DR1 statistical uncertainty, $\sigma_\mathrm{DR1}$, for each cosmological parameter is obtained by sampling posteriors using the mean of the mock power spectra (including the $\theta$-cut), with the corresponding FFA EZmock covariance matrix. 

\subsection{External Mock Data}
\label{sec:externaldata}
Due to strong projection effects, DESI DR1 data alone, even when including BAO reconstruction, does not effectively constrain parameters in extended models such as $w_0w_a$CDM without the incorporation of additional external datasets or highly informative priors~\citep{DESI2024.V.KP5}. Type Ia supernovae (SNe Ia) function as standardisable candles, providing an alternative way to measure the expansion history of the universe and can help break parameter degeneracies, thereby mitigating projection effects. To ensure robust constraints, we supplement the DESI full-shape analysis with a SNe Ia mock data set when studying $w_0w_a$CDM. We construct a Pantheon+~\citep{Scolnic2022} -- like mock data set following the publicly available likelihood from~\citep{Brout2022}, based on an underlying cosmology consistent with the \texttt{baseline} cosmology. The apparent magnitudes, denoted by $m_b$, are computed from the theoretical luminosity distances in the \texttt{baseline} cosmology, assuming a fixed absolute magnitude of $M_b = - 19.263$. 
To account for observational uncertainties, we incorporate correlated noise derived from the Pantheon+ covariance matrix. This approach ensures that our mock dataset preserves the statistical properties of real Pantheon+ -- like SNe observations while maintaining consistency with the fiducial cosmology of the `complete' Abacus-2 mocks. 
Similarly to Section~\ref{sec:DR1error}, we define the combined statistical uncertainty from DESI DR1 and the SNe Ia mock in the context of $w_0w_a$CDM as $\sigma_\mathrm{DR1+SN}$. The resulting numerical values of $\sigma_\mathrm{DR1}$ and $\sigma_\mathrm{DR1+SN}$ used throughout this work are reported in Appendix~\ref{app:appendix}.

\begin{figure}[htb]
    \centering
    \includegraphics[width=\linewidth]{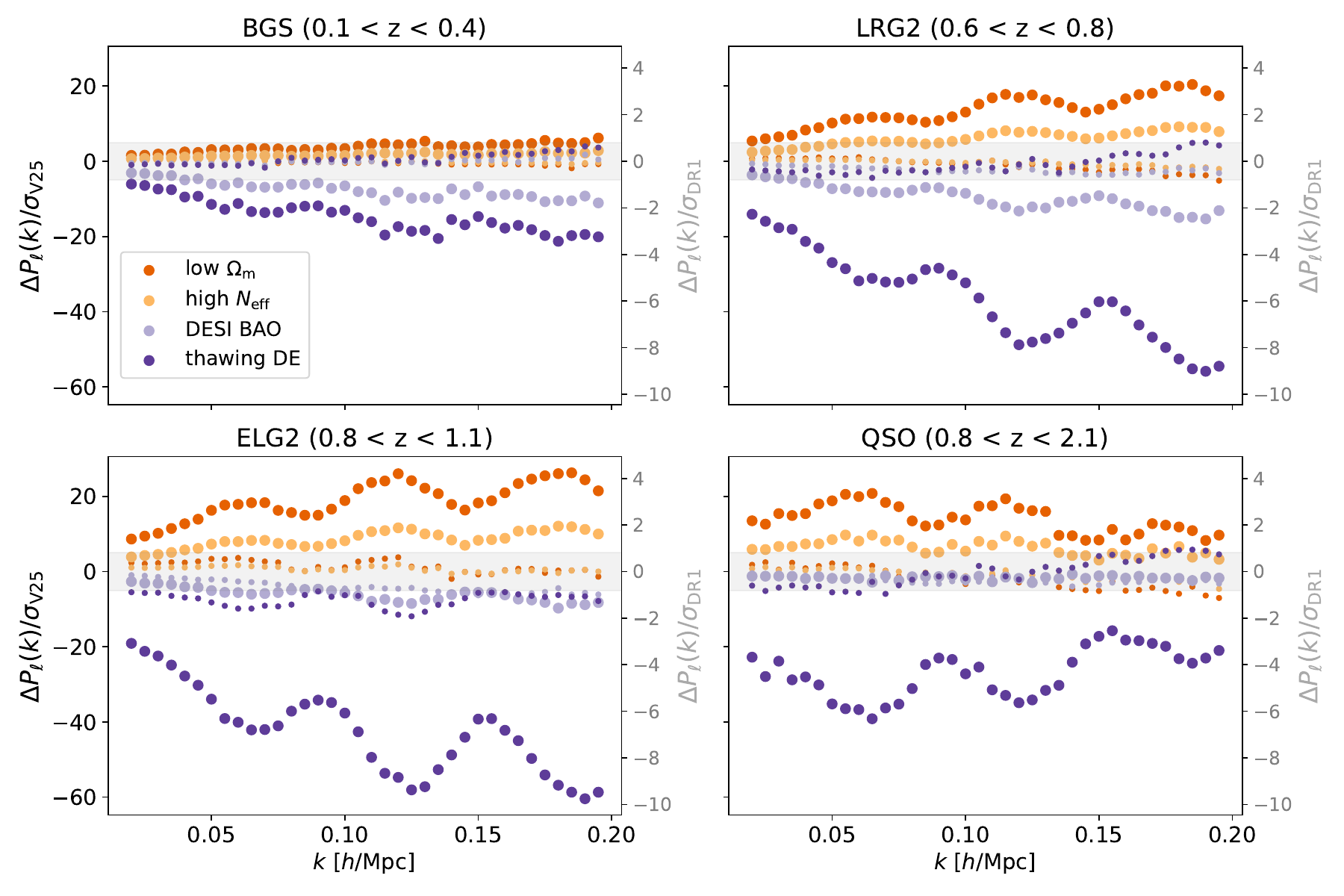}
    \caption{Residuals of the monopole (large dots) and quadrupole (small dots) power spectrum measurements for four distinct redshift bins and tracers, comparing the mean of 25 AbacusSummit mock realisations under a secondary \textit{grid} cosmology (\texttt{low-$\Omega_m$},  \texttt{high-$N_\mathrm{eff}$}, \texttt{DESI BAO},\texttt{thawing-DE}) to the DESI $\texttt{baseline}$ choice. The residuals are normalised by their combined $1\sigma_\mathrm{V25}$ uncertainities of the measurements (left axis), with a corresponding interpretation in terms of the DESI DR1 error $\sigma_\mathrm{DR1}$ shown on the right axis. The grey band indicates the $5\sigma_\mathrm{V25}$ boundary.}
    \label{fig:data}
\end{figure}

\section{Methods}
\label{sec:Method}

\begin{table}[htb]
    \centering
    \begin{tabular}{|c|c|}
        \hline
        Compressed parameters (SF) & Priors \\\hline
                $\alpha_{\parallel}$ &  $\mathcal{U}$[0.8, 1.2] \\
                $\alpha_{\perp}$ &  $\mathcal{U}$[0.8, 1.2] \\
                $f/f_{\rm fid}$ & $\mathcal{U}$[0.0, 2.0] \\
                $m$ & $\mathcal{U}$[-0.8, 0.8] \\
\hline\hline
        Cosmological parameters (FM) & Priors \\\hline
        $\omega_\mathrm{cdm}$ & $\mathcal{U}$[0.01,0.99] \\
        $\omega_\mathrm{b}$ & $\mathcal{N}[0.02237,0.00055^2]$ \\
        $h$ & $\mathcal{U}$[0.2,1] \\
        ln($10^{10}A_s$) & $\mathcal{U}$[1.61,3.91] \\
        $n_s$ & $\mathcal{N}[0.9649,0.042^2]$ \\
        $w_0$ & $\mathcal{U}$[-3.0,1.0] \\
        $w_a$ & $\mathcal{U}$[-3.0,2.0] \\
        \hline \hline
        Nuisance parameters & Priors \\
        \hline
        $(1+b_1) \sigma_8$ & $\mathcal{U}$[0,3] \\
        $b_2 \sigma_8^2$ & $\mathcal{N}[0,5^2]$ \\
        $b_s \sigma_8^2$ & $\mathcal{N}[0,5^2]$ \\
        $\alpha_0$ & $\mathcal{N}[0,12.5^2]$ \\
        $\alpha_2$ & $\mathcal{N}[0,12.5^2]$ \\
        SN$_0$ & $\mathcal{N}[0,2^2] \times 1/\bar{n}_g$ \\
        SN$_2$ & $\mathcal{N}[0,5^2] \times f_{\rm sat} \sigma_{1\,\rm eff}^2/\bar{n}_g$ \\
        \hline
    \end{tabular}
    \caption{
    Priors on cosmological and nuisance parameters employed in the full-shape analysis of this paper, following the baseline choices of the DR1 galaxy full-shape analysis in~\citep{DESI2024.V.KP5}. The only exception is the prior on $\omega_b$, which retains the same standard deviation but is centered on the true value from the Abacus-2 DR1 mocks. $\mathcal{U}$ denotes a uniform prior, while $\mathcal{N}(\mu,\sigma^2)$ represents a Gaussian prior with mean $\mu$ and standard deviation $\sigma$. The bias parameters $b_1,b_2,b_s$ are defined in the Lagrangian basis, and nuisance parameter priors follow a “physically motivated” parametrisation as described in~\citep{KP5s2-Maus}. Within this framework, counterterms scale relative to the linear theory multipoles, while stochastic terms scale with the Poissonian shot noise, $1/\Bar{n}_g$, and the characteristic halo velocity dispersion, $f_{\rm sat} \sigma_{1\,\rm eff}^2/\bar{n}_g$, where $f_{\rm sat}$ and $\sigma_{1\,\rm eff}$ represent the expected fraction and mean velocity dispersion of satellite galaxies, respectively. } 
    \label{tab:priors}
\end{table}

We present an estimate for the systematic error introduced by the choice of fiducial cosmology, based on results from the ‘complete’ Abacus-2 DR1 mock catalogues. Our analysis follows the standard full-shape pipeline, in accordance with the fiducial settings established by the DESI collaboration~\citep{DESI2024.V.KP5} (unless explicitly stated otherwise). The main objective is to assess a range of fiducial cosmologies in order to identify any systematic differences in the measured compressed or cosmological parameters relative to their expected values. Investigations into the impact of fiducial cosmology assumptions generally follow two approaches or a combination thereof. The first approach involves performing tests with a set of mocks generated from a single underlying true cosmology, where the reference cosmology of the pipeline is systematically varied. The second approach tests a set of mocks produced from different underlying true cosmologies and analysing them using a fixed-fiducial-cosmology pipeline. In this study, we adopt the first approach, systematically varying the reference cosmology throughout the entire pipeline and performing full-shape fits on the 25 realisations of the Abacus-2 DR1 mocks, whose underlying true cosmology coincides with \texttt{baseline}. Specifically, we examine five secondary pipeline cosmologies—\texttt{low-$\Omega_m$}, \texttt{thawing-DE}, \texttt{high-$N_\mathrm{eff}$}, \texttt{low-$\sigma_8$}, and \texttt{DESI BAO}—and compare them to the \texttt{baseline} choice.

\subsection{2-pt Clustering Measurements}
The clustering measurements introduced in Section~\ref{sec:AbacusSummit} were performed
separately for the South Galactic Cap (SGC) and North Galactic Cap (NGC) regions before combining them through a weighted average.  In Fourier space, this is done by averaging the power spectra of the two regions, weighted by their respective effective volumes (see ~\citep{DESI2024.II.KP3} for
more details). Our analysis follows the redshift binning scheme of table 1 in~\citep{DESI2024.V.KP5}, covering the following galaxy samples and redshift ranges: BGS ($0.1 < z < 0.4$), LRG1 ($0.4 < z < 0.6$), LRG2 ($0.6 < z < 0.8$), LRG3 ($0.8 < z < 1.1$), ELG2 ($1.1 < z < 1.6$), and QSOs ($0.8 < z < 2.1$). To represent the angular dependence with respect to the line of sight, the data was projected onto Legendre multipoles. In accordance with the baseline choices of~\citep{DESI2024.V.KP5}, we employ the power spectrum measurement for the monopole and quadrupole only, while excluding the hexadecapole to reduce prior-weight effects in the analysis. Fig.~\ref{fig:data} illustrates the impact of changing the \textit{grid} cosmology on the power spectrum measurements by presenting residuals of the monopole and quadrupole relative to the \texttt{baseline} choice across four representative redshift bins.\\

\subsection{Theoretical Modelling}
To model the non-linear redshift-space power spectrum, we transform the linear power spectrum computed using the Einstein–Boltzmann code \texttt{CLASS}\footnote{\url{https://github.com/lesgourg/class_public}}~\citep{class_2011} within the framework of EFTofLSS. The DESI collaboration has tested and compared four different Fourier space EFTofLSS-based perturbation codes~\citep{KP5s1-Maus,KP5s2-Maus,KP5s3-Noriega,KP5s4-Lai} and a configuration space EFTofLSS-based perturbation code~\citep{KP5s5-Ramirez}, all of which show consistent performance. For this analysis, we adopt the Fourier space code \texttt{velocileptors}\footnote{\url{https://github.com/sfschen/velocileptors}}~\citep{KP5s2-Maus}, which models the redshift-space power spectrum using one-loop Lagrangian Perturbation Theory (LPT) with an IR resummation scheme up to a scale $k_\mathrm{IR}$~\citep{2021Chen}. The galaxy bias model of \texttt{velocileptors} introduces four galaxy bias parameters: $b_1$ and $b_2$, which characterise the linear and non-linear biases, respectively; $b_s$, which describes the non-local tidal bias; and $b_3$, which accounts for third-order non-linear bias contributions. Following the DESI full-shape analysis~\citep{DESI2024.V.KP5}, we set the $b_3$ bias parameter to zero due to its degeneracy with the counterterms. We then use the Eulerian Perturbation Theory (EPT) module of \texttt{veloclileptors}, which reparameterises the Lagrangian bias expansion as in~\citep{Chen_2020}, to compute the redshift-space power spectrum monopole and quadrupole. Additionally to the galaxy bias parameters, we include two stochastic parameters, $\mathrm{SN}_0$ and $\mathrm{SN}_2$, along with two counterterm parameters, $\alpha_0$ and $\alpha_2$, where the subscripts denote their respective contributions to the monopole and quadrupole. 

\subsection{Cosmological Inference}
For the FM approach, we consider sampling two cosmological models: (i) a standard \lcdm{} model, which we assume when initially analysing all the secondary fiducial cosmology cases; and (ii) a beyond-\lcdm{} parameterisation incorporating $w_0$ and $w_a$, used for the \texttt{thawing-DE} and \texttt{DESI BAO} scenario. In the compressed approach, we employ the SF parameterisation, consisting of $\{ \alpha_\parallel, \alpha_\perp, m, n, f\sigma_{s8} \}$. We fix the template amplitude and fit for $f$, where $f$ can be reinterpreted as $f\sigma_{s8}$ by multiplying it with the fixed $\sigma_{s8}^\mathrm{fid}$. This holds exactly at first-order perturbation theory and remains a good approximation for higher-order corrections.
It is possible to convert $f\sigma_{s8}$ to $f\sigma_{8}$ through a scale rescaling via ${q}_\mathrm{iso}$, as detailed in Eq. (3.5) of~\citep{Brieden_2021JCAP}. In this analysis, we vary only $m$, keeping $n$ fixed, so that in the final interpretation step, $m$ effectively represents $m + n$. 

For cosmological inference in both SF and FM approaches, we impose flat priors on all cosmological and compressed parameters, except for $n_s$ and $\omega_b$ in the FM case. In the FM analysis, we adopt a Gaussian Big Bang Nucleosynthesis-like prior on the true $\omega_b$ value of the mocks, setting it to $\omega_b = 0.02237\pm0.00055$~\citep{Schoeneberg_2024} and a broad Gaussian prior on $n_s$ with a width ten times the $1\sigma$ Planck error~\citep{Planck2018}.  Table~\ref{tab:priors} summarises the priors applied to SF, FM, and nuisance parameters. Given the weak constraining power of the DESI full-shape data on $\omega_b$ and $n_s$, we omit results for these parameters. We assess the impact of the fiducial cosmology by examining shifts in the MAP values under flat priors on bias and nuisance terms. The model is fitted to the monopole and quadrupole over $k = 0.02 - 0.20~h{\rm Mpc}^{-1}$ using the \texttt{Minuit}\footnote{\url{https://github.com/scikit-hep/iminuit}} profiler~\citep{minuit} to determine MAP values. For the purpose of beyond-\lcdm{} parameterisation in the FM approach, Markov Chain Monte Carlo (MCMC) sampling is not employed when evaluating the fiducial cosmology contribution. This is due to the fact that projection effects can impact the posterior mean but not the MAP estimate in extended models (see~\citep{DESI2024.V.KP5} for discussion). However, when comparing MAP values to the marginalised posterior in figures, we employ the Metropolis-Hastings
MCMC sampler~\citep{Lewis2002,Lewis2013} within \texttt{cobaya}~\citep{Torrado2021}\footnote{\url{https://cobaya.readthedocs.io/en/latest/index.html}}. In this case, the linear nuisance parameters, namely $\alpha$ and SN, are analytically marginalised to accelerate the sampling process. All modelling and inference routines are implemented within the DESI likelihood pipeline \texttt{desilike}\footnote{\url{https://github.com/cosmodesi/desilike}}. To improve computational efficiency, we employ a fourth-order Taylor expansion to emulate the theory model when sampling compressed or \lcdm{} parameters. This approach takes advantage of the fact that the predicted power spectrum multipoles are a smooth function of the underlying cosmological parameters reasonably close to their chosen fiducial values~\citep{Colas2020,Chen2022}. The emulator was tested in~\citep{KP5s2-Maus} and showed good agreement with the direct model predictions when expanded to fourth order. However, we do not employ the emulator in the FM analysis of $w_0w_a$CDM, where deviations from the fiducial model could render the emulator inaccurate.

\subsection{Definition of Shift Parameters}
\label{sec:shiftparameters}
In this section, we introduce the shift parameters used to quantify potential biases in the recovered cosmological or compressed parameters after accounting for the expected effects of the fiducial cosmology. These shift parameters allow us to assess the robustness of our analysis against systematic uncertainties arising from assumptions about the underlying cosmological model.
For any given cosmological or compressed parameter, we define the shift relative to its expected value under a chosen fiducial cosmology. Specifically, we define:
\begin{align}
\delta x = x^\mathrm{meas.} - x^\mathrm{exp.},
\end{align}
where $x^\mathrm{meas.}$ denotes the parameter recovered from the data, and $x^\mathrm{exp.}$ corresponds to the expected value assuming the DESI \texttt{baseline} cosmology is the true underlying cosmology. 
By analysing these shifts, we can diagnose whether our inference introduces systematic biases when different fiducial cosmologies are used. In an ideal scenario where the analysis is unbiased, we expect these shifts to be small and consistent with statistical fluctuations of the $V_{25}$ volume. To mitigate sample variance, we focus on the difference between the shifts in the secondary cosmologies (\texttt{low-$\Omega_m$}, \texttt{thawing-DE}, \texttt{high-$N_\mathrm{eff}$}, \texttt{low-$\sigma_8$}, and \texttt{DESI BAO}) with respect to the $\texttt{baseline}$ pipeline choice:
\begin{align}
\Delta x = \delta x^\mathrm{secondary~cosmo.} - \delta x^\texttt{baseline}.
\label{eq:shift}
\end{align}
Since both $\delta x^\mathrm{secondary~cosmo.}$ and $\delta x^\texttt{baseline}$ are computed from the same set of mocks, they share a common realisation of cosmic variance. By taking their difference, we cancel out part of the sample variance that affects both measurements in a correlated way. This allows us to isolate systematic effects introduced by changes in the fiducial cosmology while reducing statistical noise. Any statistically significant deviation of $\Delta x$ from zero would then indicate the presence of a systematic bias. \\

\section{Results}
\label{sec:Results}
In this section, we present the results of the fiducial cosmology dependence of the DESI full-shape analysis, which is used to establish the final systematic error budget in~\citep{DESI2024.V.KP5}. These results are derived from the Abacus-2 DR1 mocks (with an underlying true cosmology equal to \texttt{baseline})  and are analysed using both the FM and SF approaches.

\begin{figure}[htb]
    \centering
    \includegraphics[width=0.95\linewidth]{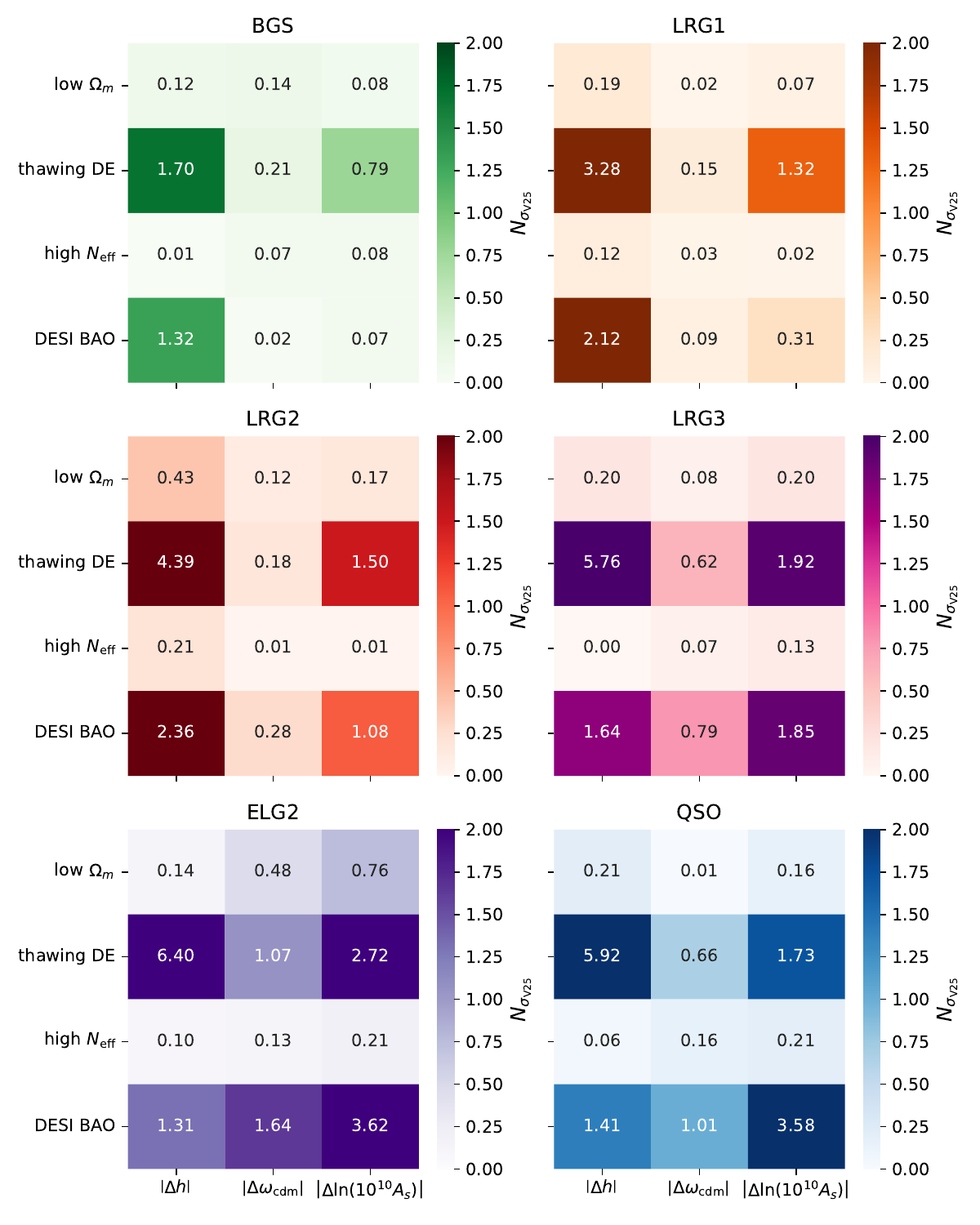}
    \caption{Heatmap of the shifts in cosmological parameters relative to the \texttt{baseline} cosmology, expressed in units of $N_{\sigma_\mathrm{V25}}$, as derived from FM fits to the mean of 25 Abacus-2 DR1 complete mock realisations for each tracer.  While for \texttt{low-$\Omega_m$} and \texttt{high-$N_\mathrm{ eff}$}, all shifts remain below $2\sigma_\mathrm{V25}$, \texttt{thawing-DE} and \texttt{DESI BAO} exhibit deviations up to $6.40\sigma_\mathrm{V25}$. This 
    highlights the inability of a restricted \lcdm{} model to fully correct distortions introduced by a beyond \lcdm{} \textit{grid} cosmology.
    }

    \label{fig:FM_AS_heatmap}
\end{figure}

\subsection{Full-modelling Results}
\label{sec:FMresults}
In order to systematically analyse the impact of different \textit{grid} cosmologies, we consider two groups when performing FM analyses, depending on the imposed sampled cosmological model. In the first group, we assume a \lcdm{} framework, where only standard \lcdm{} parameters are varied. Within this group, we test all four secondary \textit{grid} cosmologies. The second group is treated within an extended $w_0w_a$CDM cosmological context, where additional parameters ($w_0, w_a$) beyond \lcdm{} are sampled. In this group, we only consider the extended \textit{grid} cosmologies \texttt{thawing-DE} and \texttt{DESI BAO}. 
To fully understand the impact of different model assumptions, we analyse the extended \textit{grid} cosmologies using both \lcdm{} and $w_0w_a$CDM parameterisations. The \lcdm{} inference in this context serves as a stress test, illustrating the limitations of interpreting beyond \lcdm{} \textit{grid} cosmologies within a restricted framework. In contrast, the $w_0w_a$CDM inference allows for greater flexibility, ensuring that any distortions introduced by an extended \textit{grid} cosmology are properly accounted for by varying the corresponding parameters during the inference process.  We do not employ the same approach for \texttt{high-$N_\mathrm{eff}$} because, although an increased effective number of relativistic species introduces additional degrees of freedom, its impact is primarily on the early-time evolution of the Universe. This results in shifts in the derived parameters rather than a fundamental extension of the late-time parameter space, as it is the case for \texttt{thawing-DE} or \texttt{DESI BAO}.

\begin{figure}[htb]
    \centering
    \includegraphics[width=0.85\linewidth]{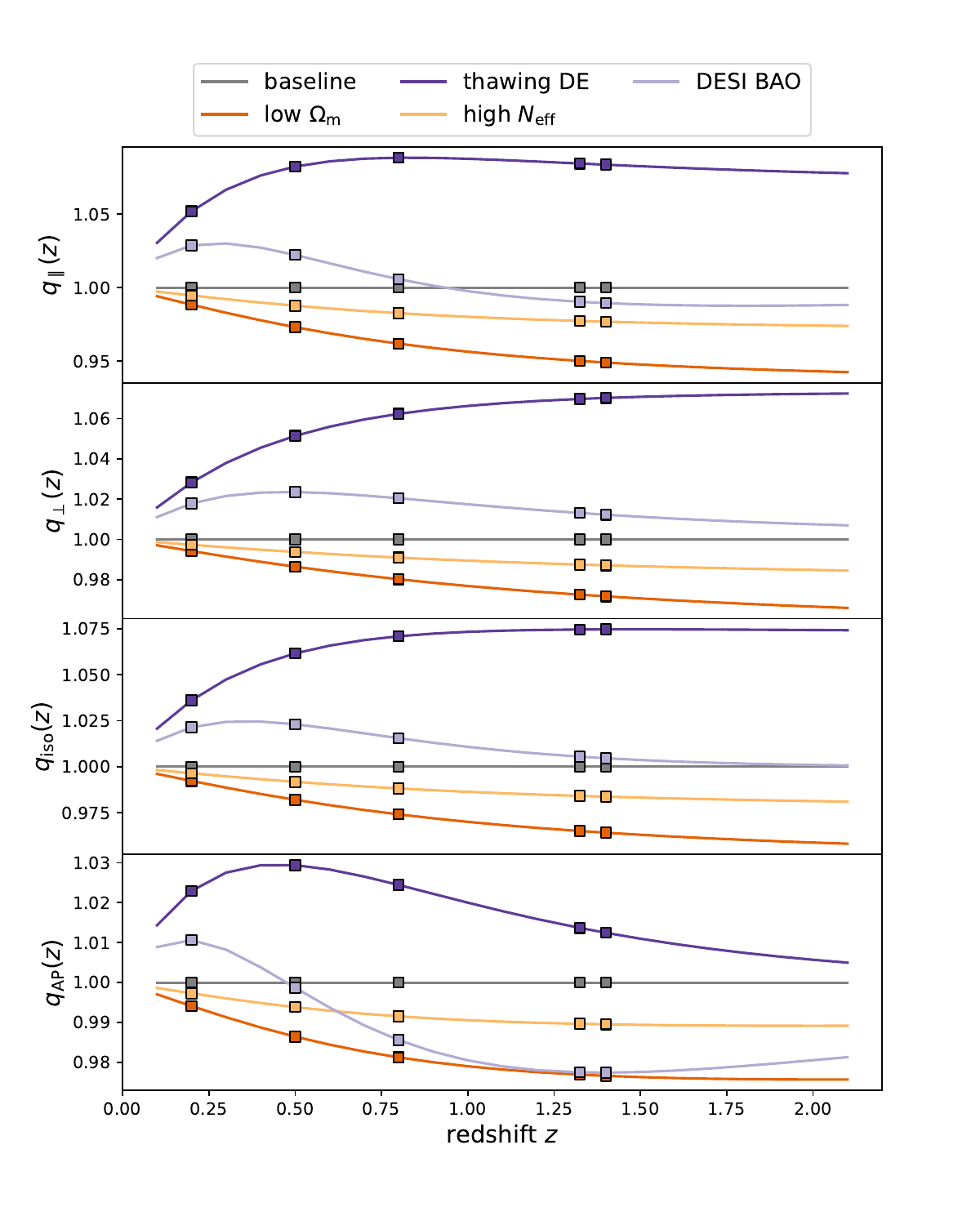}
    \caption{Scaling parameters as a function of redshift for different \textit{grid} cosmologies, showing the geometric distortions introduced when assuming the wrong cosmology for the redshift-to-distance conversion. The parameters $q_\parallel(z)$ and $q_\perp(z)$ quantify the line-of-sight and the transverse dilation effects, respectively, while 
    $q_\mathrm{iso} (z)  =[ q_\parallel(z) q_\perp(z)^2]^{1/3} $ and $q_\mathrm{AP} = q_\parallel(z)/q_\perp(z)$ capture isotropic and anisotropic combinations. Squares indicate the effective redshifts of the six DESI tracers: BGS, LRG1, LRG2, LRG3, ELG2, and QSO.}
    \label{fig:scaling_params}
\end{figure}

\subsubsection{\texorpdfstring{\lcdm}{LambdaCDM}}
Figure~\ref{fig:FM_AS_heatmap} presents the shifts in cosmological parameters relative to the DESI \texttt{baseline} cosmology, computed using the MAP values from fits to the mean of 25 Abacus-2 DR-1 complete mock realisations for each tracer. These shifts are expressed in terms of
\begin{equation}
    N_{\sigma_\mathrm{V25}} \equiv \frac{\left|\Delta x\right|}{\sqrt{\sigma^2_{\mathrm{Abacus}, \texttt{baseline}}(x)+ \sigma^2_{\mathrm{Abacus}, \mathrm{secondary~cosmology}}}(x)}, 
    \label{eq:shiftV25}
\end{equation}
where $\Delta x$ represents the difference in shifts between the secondary cosmology and the \texttt{baseline} as defined in Eq.~\eqref{eq:shift}. We consider a shift as statistically significant if it exceeds $2\sigma_\mathrm{V25}$, given the precision of the 25 realisations.

\begin{figure}[htb]
    \centering
    \includegraphics[width=0.85\linewidth]{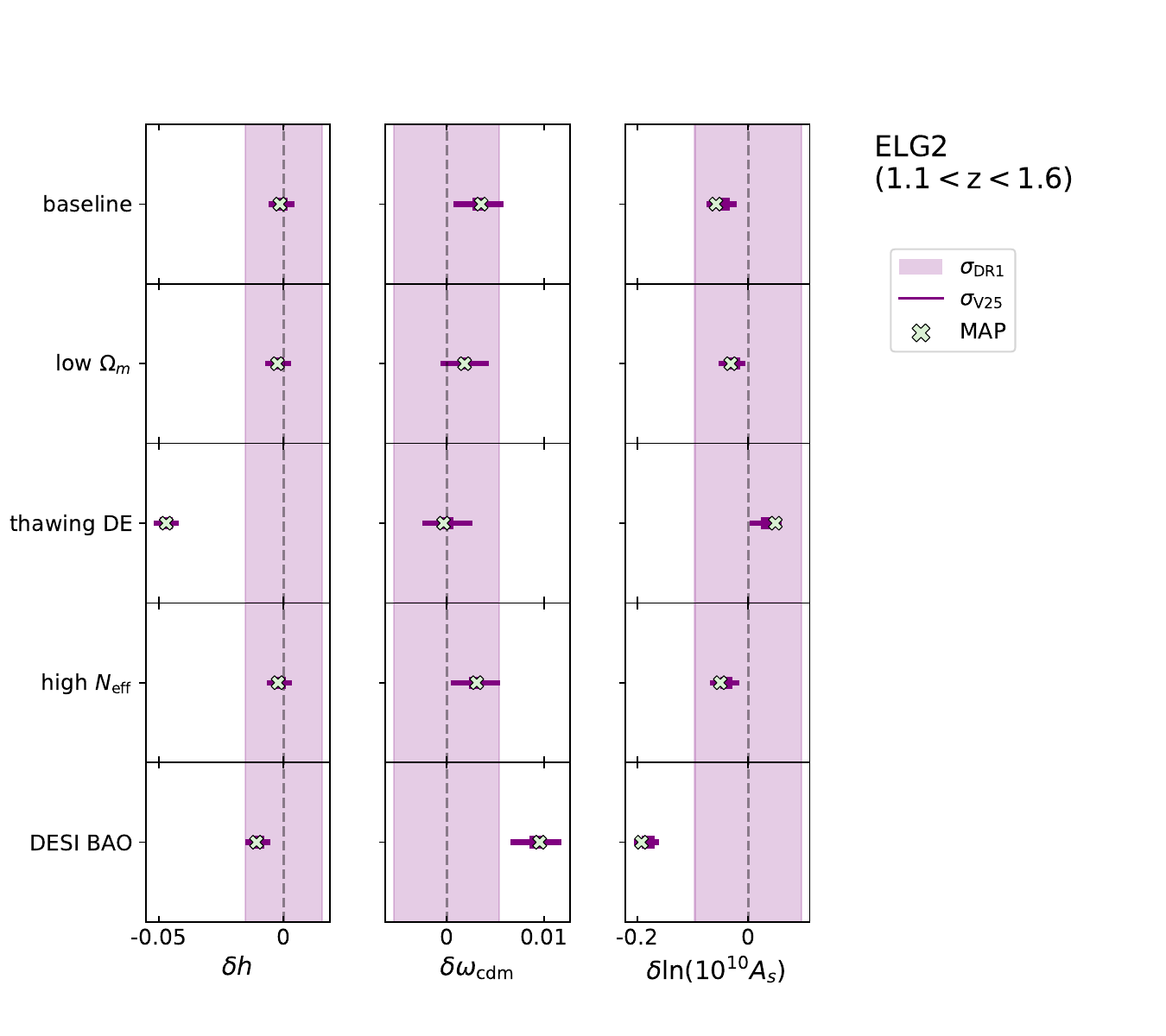}
    \caption{1D marginalised posterior distributions for the ELG2 sample under the assumption of different fiducial cosmologies:  \texttt{baseline, low-$\Omega_m$, thawing-DE, high-$N_\mathrm{eff}$, DESI BAO}. The purple squares indicate the mean of the marginalised posteriors, with dark horizontal lines representing the corresponding $1\sigma_\mathrm{V25}$ error bars derived from fits to the mean of 25 Abacus-2 DR1 complete mock realisations. The purple shaded region shows the expected realistic DR1 statistical uncertainty, centered on a zero shift with respect to the true cosmological parameters. Green crosses mark the MAP values. 
    }
    \label{fig:FM_ELG}
\end{figure}

While the shifts are below $2\sigma_\mathrm{V25}$ for all tracers and parameters in the \lcdm{}-like secondary cosmologies (\texttt{low-$\Omega_m$} and \texttt{high-$N_\mathrm{eff}$}), the shifts are becoming more pronounced at higher redshifts. This trend reflects the greater sensitivity of the inferred clustering parameters to the assumed \textit{grid} cosmology at high redshifts, where errors in the redshift-to-distance conversion accumulate over larger cosmic volumes. The redshift-dependent impact of the \textit{grid} cosmology can also be seen in the power spectrum measurement shown in Figure~\ref{fig:data}, where the residual differences between the \texttt{baseline} cosmology and the secondary cosmologies are displayed. Among the two \lcdm{}-like secondary cosmologies, the $\texttt{low-$\Omega_\mathrm{m}$}$ case produces the largest shifts, particularly in the highest redshift bins of the LRG and ELG samples. However, even these shifts remain below $2\sigma_\mathrm{V25}$, indicating that they are fully consistent with statistical fluctuations. Therefore, we conclude that adopting an alternative \lcdm{} fiducial cosmology has no measurable effect on the inferred parameters. The larger shifts observed in the $\texttt{low-$\Omega_\mathrm{m}$}$ case are expected, as $\Omega_\mathrm{m}$ directly affects the late-time expansion history and, consequently, the redshift-to-distance relationship. In contrast, the \texttt{high-$N_\mathrm{eff}$} scenario primarily affects the sound horizon scale $r_\mathrm{d}$ and has therefore a stronger impact on template-based analysis (see Section~\ref{sec:SFresults} or~\citep{KP4s9-Perez-Fernandez}) than FM approaches. \\

For completeness, we now extend this analysis to include the remaining two \textit{grid} cosmologies: \texttt{thawing-DE} and \texttt{DESI BAO}, both of which deviate considerably from the \texttt{baseline} \lcdm{} expansion history. Figure~\ref{fig:scaling_params} shows the corresponding scaling parameters $q_\perp(z)$ and $q_\parallel(z)$ (see Eqs.~\eqref{eq:AP_FM1} \&~\eqref{eq:AP_FM2}), as well as the derived isotropic and anisotropic combinations $q_\mathrm{iso} (z)  =[ q_\parallel(z) q_\perp(z)^2]^{1/3} $ and $q_\mathrm{AP} = q_\parallel(z)/q_\perp(z)$, under the assumption of the \texttt{baseline} cosmology as the true cosmology. The squares represent the effective redshifts for each tracer: BGS ($z_\mathrm{eff} = 0.2$), LRG1 ($z_\mathrm{eff} = 0.5$), LRG2 ($z_\mathrm{eff} = 0.8$), LRG3 ($z_\mathrm{eff} = 0.8$), ELG2 ($z_\mathrm{eff} = 1.325$) and QSO ($z_\mathrm{eff} = 1.4$). Among all \textit{grid} cosmologies considered, the \texttt{thawing-DE} case leads to the strongest geometric distortions, reaching deviations up to $7-9\%$ in both $q_\parallel$ and $q_\perp$. The \texttt{DESI BAO} case reflects the distinct phantom-like behaviour exhibited by the DESI best-fit $w_0w_a$CDM cosmology, with $q_\parallel(z)$ crossing from below unity at high redshift to above unity at low redshift. Figure~\ref{fig:FM_AS_heatmap} shows that interpreting these beyond \lcdm{} \textit{grid} cosmologies using a restricted sampled \lcdm{} model results in significant shifts, with values reaching up to $6.4\sigma_\mathrm{V25}$ in the context of \texttt{thawing-DE} and $3.6\sigma_\mathrm{V25}$ for the \texttt{DESI BAO} scenario. 
The existence of these biases can be understood through two complementary mechanisms, both driven by the inability of a restricted \lcdm{} parameter space to accommodate the distortions introduced by a beyond-\lcdm{} grid cosmology. The first mechanism is specific to the FM approach and explains why the shifts seen here are generally more pronounced than those observed in the SF approach (see Section~\ref{sec:SFresults}). In the SF framework, the compressed parameters $\{ \alpha_\parallel, \alpha_\perp, m, n, f\sigma_{s8} \}$ allow sufficient freedom to absorb much of the geometric distortion introduced by an incorrect grid cosmology, since the assumed cosmological model enters only at the final interpretation step. In the FM approach, however, the cosmological parameters are directly sampled within a restricted \lcdm{} space, and the inability of this model to reproduce the distortions induced by a beyond-\lcdm{} grid cosmology leads to strong parameter degeneracies — most notably between $h$, $\ln(10^{10} A_s)$, and the bias parameter $b_1$ — resulting in biased MAP estimates. The second mechanism affects both FM and SF, and arises from the limitations of the geometric correction to only account for linear distortions. In contrast, the effect of measuring the full power spectrum under the assumption of a beyond \lcdm{} cosmology also alters the shape and scale-dependence, which can not be undone by a simple rescaling. Additionally, these distortions are redshift-dependent, and a single pair of rescaling parameters evaluated at an effective redshift is often insufficient to fully capture the impact within a redshift bin. As a result, even if $D_M^\mathrm{grid}(z_\mathrm{eff})$ and $D_H^\mathrm{grid}(z_\mathrm{eff})$ are correctly specified, the residual evolution across a redshift bin can introduce biases in the inferred cosmological parameters, especially if the scaling parameters vary rapidly, as is the case for $q_\parallel(z)$ for \texttt{DESI BAO}. Figure~\ref{fig:FM_ELG} illustrates this effect by showing the 1D marginalised posterior distributions for the ELG tracer — the tracer that exhibits the largest shifts in Figure~\ref{fig:FM_AS_heatmap} - under different \textit{grid} cosmologies.  The purple square symbols represent the mean of the marginalised posteriors, with horizontal error bars indicating the corresponding $1\sigma_\mathrm{V25}$ regions derived from fits to the mean of 25 Abacus-2 DR1 complete mock realisations. The purple shaded region denotes the expected realistic DR1 statistical uncertainties, centered on a zero shift with respect to the true cosmological parameters. Additionally, green crosses mark the MAP values. The \texttt{thawing-DE} \textit{grid} cosmology particularly biases the Hubble parameter $h$, reflecting the slower late-time evolution of the expansion rate. In contrast, the most significantly affected parameters under the assumption of the \texttt{DESI BAO} \textit{grid} cosmology are the amplitude of primordial fluctuations $\ln(10^{10} A_s)$, the scalar spectral index $n_s$ (not shown here), and the cold dark matter density $\omega_\mathrm{cdm}$, due to differences in the shape of the expansion history at intermediate redshifts, which affect the scale-dependent growth and the mapping of modes in $k$-space. These results confirm that interpreting beyond \lcdm{} \textit{grid} cosmologies within a restricted sampled \lcdm{} model can lead to misleading or biased inferences. This motivates the complementary $w_0w_a$CDM analysis discussed in Section~\ref{sec:beyondlcdm}, where the additional flexibility in the sampled model significantly improves the robustness of the inference for extended cosmologies. In the rest of this section, we restrict our attention to the \lcdm{}-like \textit{grid} cosmologies (\texttt{low-$\Omega_m$} and \texttt{high-$N_\mathrm{eff}$}), while systematic error contributions from the more strongly deviating \textit{grid} cosmologies (\texttt{thawing-DE} and \texttt{DESI BAO}) are not included here and will be assessed using the beyond \lcdm{} parameter fits presented in Section~\ref{sec:beyondlcdm}.\\ 

To summarise the systematic impact of the fiducial cosmology choice, the upper rows of Table~\ref{tab:max_shifts} report the maximum shifts observed in the \lcdm{} framework analysis expressed in terms of realistic DESI DR1 statistical uncertainties using
\begin{equation}
    N_{\sigma_\mathrm{DR1}} 
    \equiv \frac{\left|\Delta x\right|}{\sigma_\mathrm{DR1}},
    \label{eq:DR1-error}
\end{equation}
where $\sigma_\mathrm{DR1}$ represents the DR1 statistical error. The maximum contribution from fiducial cosmology variations remains below $0.2 \sigma_\mathrm{DR1}$ for all parameters across the individual tracers, indicating that adopting a different \lcdm-like fiducial cosmology introduces only minor systematic biases in the inferred cosmological constraints. These shifts provide an upper bound for our estimate of the systematic error budget due to \lcdm{} like fiducial cosmologies, as reported in table~7 of~\citep{DESI2024.V.KP5}. Although the individual tracer contributions remain below $0.2 \sigma_\mathrm{DR1}$, the combined constraints  -- derived from a joint fit of all tracers to the mean of the 25 Abacus-2 DR1 complete mock realisations -- show slightly larger shifts, with values of $0.22 \sigma_\mathrm{DR1}$, $0.21 \sigma_\mathrm{DR1}$, and $0.2 \sigma_\mathrm{DR1}$ for $\Delta h$, $\Delta \omega_\mathrm{cdm}$, and $\Delta \ln(10^{10} A_s)$, respectively. It is important to note that these shifts are still well within the statistical uncertainty of the DESI DR1 full-shape analysis, confirming that the systematic impact of fiducial cosmology variations is subdominant. However, as statistical precision improves in the future, it will be important to reassess these systematic uncertainties to ensure the accuracy of cosmological constraints. In order to do so, it will be key to minimise the sample variance of the N-body mocks used to perform this study, either by running a larger number of simulations, or by employing some cancelling-variance method \cite{Ding25}. Since the DESI statistical precision will increase with future data releases, the finite volume of the mocks ($200\, [h^{-1} \rm{Gpc}]^3$ in this study) will eventually become a limiting factor in determining the systematic error budget.

\begin{table}[htb]
\centering
\begin{tabular}{|lccccc|}
    \hline
    \multicolumn{6}{|c|}{$N_{\sigma_\mathrm{DR1}}$}\\
    \hline
    Tracer  & $\Delta h$ & $\Delta \omega_{\rm cdm}$ & $\Delta \ln (10^{10} A_s)$& $\Delta w_0$ & $\Delta w_a$ \\
    \hline
    BGS     & 0.04  & 0.06  & 0.04 & -- & -- \\
    LRG1  & 0.07  & 0.01  & 0.03 & -- &  --\\
    LRG2  & 0.18  & 0.06  & 0.06 & --& --\\
    LRG3  & 0.09  & 0.03  & 0.06 & -- & --\\
    ELG2  & 0.03  & 0.16  & 0.14 & -- & --\\
    QSO  & 0.06  & 0.07  & 0.05 & -- & --\\
    combined &0.22 &0.21 & 0.20 & -- & -- \\
    \hline
    \hline
    \multicolumn{6}{|c|}{$N_{\sigma_\mathrm{DR1+SN}}$}\\
    \hline
    combined &0.11 &0.11 & 0.09 & 0.12 & 0.11 \\
    \hline
\end{tabular}
    \caption{Maximum shifts in cosmological parameters for each tracer and the combination of all tracers when considering a \lcdm{} framework (upper rows), expressed in units of the DESI DR1 statistical uncertainties, $\sigma_\mathrm{DR1}$ (see Eq.~\eqref{eq:DR1-error}), as well as in units of the combined statistical uncertainty from DESI DR1 and a Pantheon$+$-like SNe Ia mock sample, $\sigma_\mathrm{DR1+SN}$ (see Eq.~\eqref{eq:DR1SNe-error}), for an extended $w_0w_a$CDM parameter space (last row). In all cases, the shifts remain well within the corresponding statistical uncertainties, indicating that the variations in fiducial cosmology have a negligible systematic impact for both standard and extended cosmological models.}
    \label{tab:max_shifts}
\end{table}

\begin{figure}[htb]
    \centering
    \includegraphics[width=1\linewidth]{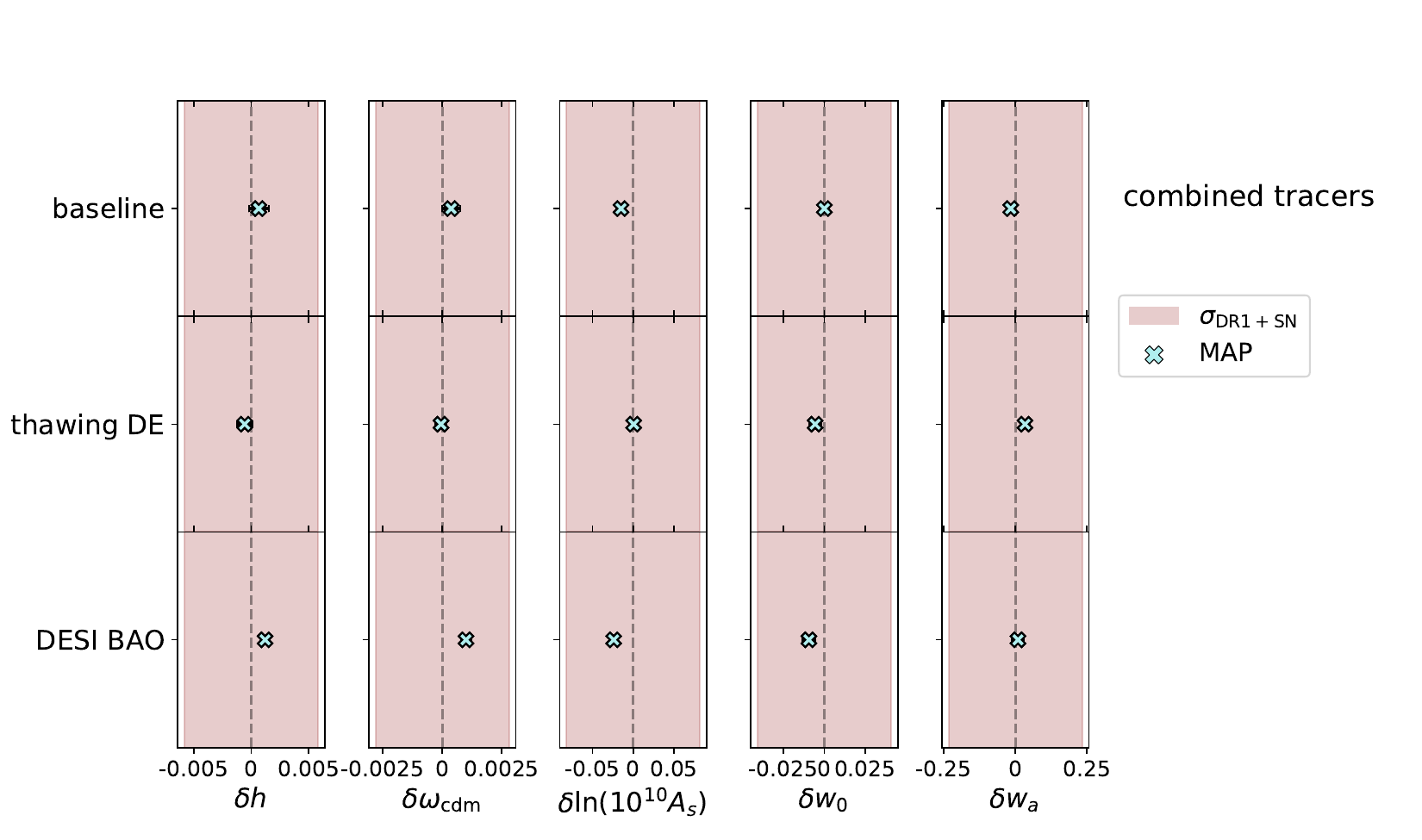}
    \caption{1D marginalised posterior distributions for the combination of the 6 DESI tracers, based on the average of the 25 Abacus-2 DR1 complete mock realisation, and a noiseless SNe Ia mock data set for $w_0w_a$CDM. Two fiducial cosmologies were studied: \texttt{baseline} and \texttt{thawing DE}.  The red shaded region shows the expected realistic uncertainties from the combination of DESI DR1 full shape results in combination with realistic SNe Ia constraints, denoted as $\sigma_{\mathrm{DR1+SN}}$ and centered on a zero shift with respect to the true cosmological parameters. Blue crosses mark the corresponding MAP values, with horizontal error bars representing the standard deviation from 10 independent minimisation runs.}
    \label{fig:FM_joint_w0wa}
\end{figure}

\subsubsection{ Beyond \texorpdfstring{\lcdm}{LambdaCDM}}
\label{sec:beyondlcdm}
In the beyond \lcdm{} group, we investigate systematic shifts in cosmological parameters due to the assumption of an extended \textit{grid} cosmology (\texttt{thawing DE} and \texttt{DESI BAO}) with respect to the DESI \texttt{baseline} choice. In this group, we vary $w_0$ and $w_a$ in addition to the standard \lcdm{} parameters during the inference process in order to accommodate the extended parameter space of the \texttt{thawing DE} and \texttt{DESI BAO} \textit{grid} cosmology as argued above. However, due to the large volume in parameter space, the DESI FM analysis alone lacks the statistical power to robustly constrain this model. Consequently, we analyse the systematic shifts in the beyond \lcdm{} group within the inclusion of external mock data sets. Specifically, we incorporate an SNe Ia mock data set (as described in Section~\ref{sec:externaldata}) as part of this study. The addition of SNe Ia data is particularly useful as it helps suppress potential projection effects arising in DESI-only analyses within marginalised parameter constraints. The combination of DESI and SNe Ia aligns with the approach taken in \citep{DESI2024.VII.KP7B}, where CMB data is included to further tighten constraints on the $w_0w_a$CDM model. 

Due to the high computational cost of generating full MCMC chains for $w_0w_a$CDM, we do not express any shifts in terms of the precision of the 25 Abacus-2 DR1 complete mock realisations for this case. Instead, we directly quantify the shifts relative to the total uncertainty of DESI DR1 
in combination with the uncertainty of a Pantheon$+$-like SNe Ia sample, $\sigma_{\mathrm{DR1+SN}}$. In analogy to Eq.~\eqref{eq:DR1-error} we report the maximum shifts in terms of the total statistical value as,
\begin{equation}
    N_{\mathrm{DR1+SN}} \equiv \frac{\left|\Delta x\right|}{\sigma_{\mathrm{DR1+SN}}},
\label{eq:DR1SNe-error}
\end{equation}
where $\Delta x$ represents the difference in shifts between the extended secondary cosmology and the \texttt{baseline} cosmology. Note that the shift, $\Delta x$, is calculated from the MAP values obtained from a joint fit of all tracers to the mean of the 25 Abacus-2 DR1 complete mock realisations combined with a {\it noiseless} SNe Ia mock data set. In this joint fit, the covariance is scaled to represent the volume of a single-volume realisation ($\rm V_1$), unlike for the $\Lambda$CDM cases, and is combined with the SNe Ia covariance with the volume of the Pantheon$+$-like SNe Ia sample. We do this to preserve a realistic relative sample weighting between the Pantheon$+$-like SNe Ia and the DESI DR1 full-shape data when determining the MAP values and errors. Figure~\ref{fig:FM_joint_w0wa} shows the 1D marginalised posterior distribution for the \texttt{baseline}, \texttt{thawing DE} and \texttt{DESI BAO} \textit{grid} cosmologies obtained from the combination of DESI and noiseless SNe Ia mocks. The red shaded region denotes the expected uncertainties ($\sigma_{\mathrm{DR1+SN}}$) for a $w_0w_a$CDM model based on DESI+SNe Ia. These uncertainties are centered on zero shift relative to the true cosmological parameters, and the corresponding MAP values are marked as blue crosses in the figure. The associated horizontal error bars reflect the standard deviation across 10 independent minimisation runs and serve as a rough estimate of the fit stability. From this analysis, we find that the change of a \lcdm{}-like fiducial cosmology to an extended $w_0w_a$CDM cosmology induces the following systematic shifts in key cosmological parameters: $0.11 \sigma_{\mathrm{DR1+SN}}$, $0.11 \sigma_{\mathrm{DR1+SN}}$, $0.09 \sigma_{\mathrm{DR1+SN}}$, $0.12 \sigma_{\mathrm{DR1+SN}}$ and $0.11 \sigma_{\mathrm{DR1+SN}}$ for $\Delta h$, $\Delta \omega_\mathrm{cdm}$, and $\Delta \ln(10^{10} A_s)$, $\Delta w_0$, and $\Delta w_a$, respectively. These results are summarised in the bottom row of Table~\ref{tab:max_shifts}. All reported shifts remain well within the total statistical uncertainty of the DESI DR1 + SNe Ia combination. These findings indicate that the choice of fiducial cosmology introduces a negligible systematic effect in the context of beyond \lcdm{} analysis when full-shape data is combined with external data at the current precision level. While combining DESI full-shape data with additional datasets reduces the overall uncertainty on parameter estimates -- potentially making any fixed systematic contribution relatively more significant -- these additional datasets also help mitigate the impact of fiducial cosmology through their complementary constraining power, which helps to further break parameter degeneracies in extended models. As a result, although the relative importance of the systematic may grow, its influence on the final parameter constraints can still be reduced when the parameter space is more tightly constrained by additional probes, such as the reconstructed DESI BAO and Planck CMB, as it is done in~\citep{DESI2024.VII.KP7B}.

\begin{figure}[htb]
    \includegraphics[width=0.9\linewidth]{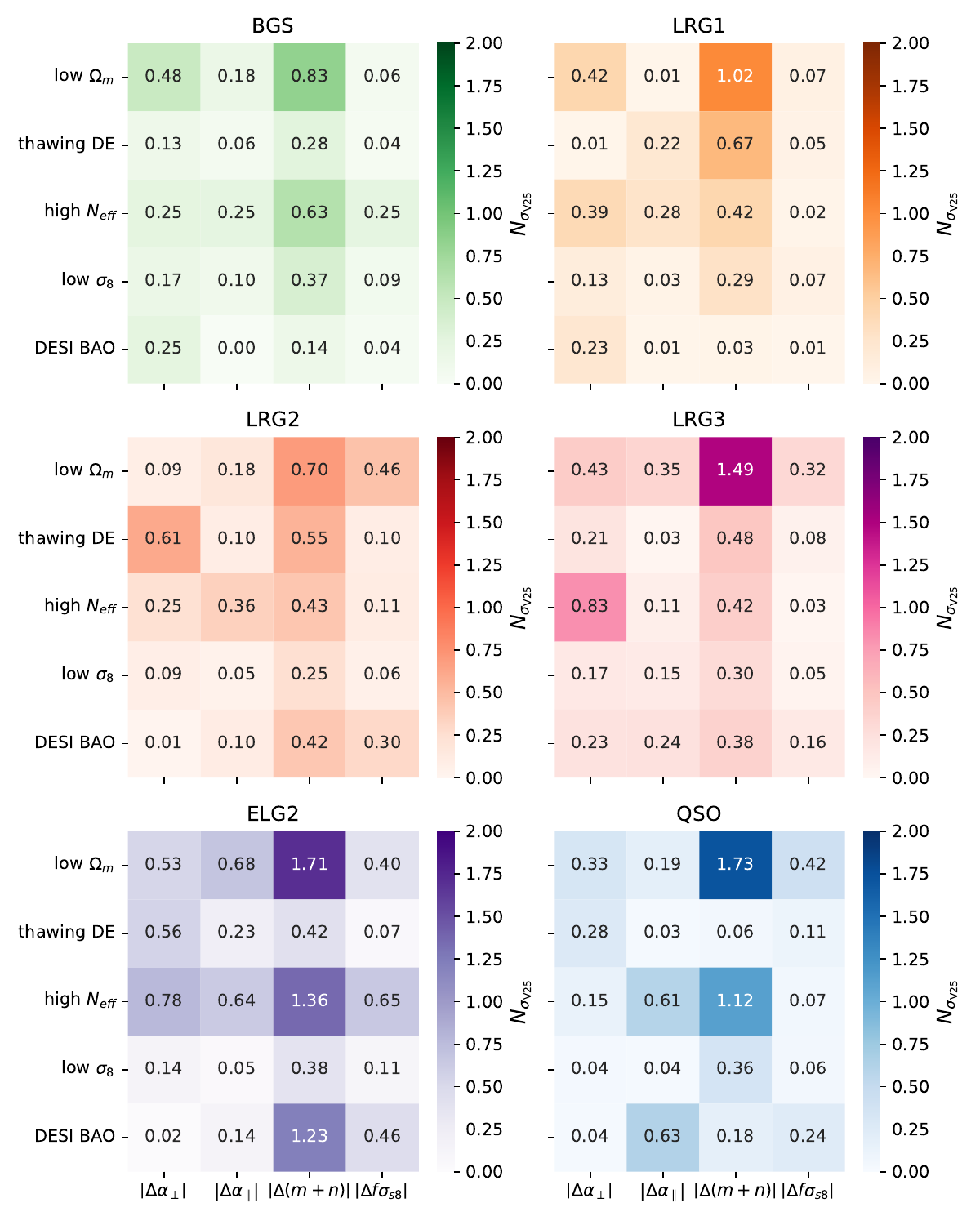}
    \caption{Same as Figure~\ref{fig:FM_AS_heatmap} but for the SF approach. The heatmap shows the shifts in cosmological parameters relative to the \texttt{baseline} cosmology in terms of the error for the volume of 25 realisations $N_{\sigma_{\mathrm{V25}}}$, across all tracers, parameters and cosmologies considered. All shifts are below 1.73$\sigma$ (relative to the volume of 25 Abacus-2 DR1 realizations), and we therefore conclude that no statistically significant systematic shifts are present in any of the parameters due to the choice of fiducial cosmology.}
    \label{fig:sf_AS_heatmap}
\end{figure}

\subsection{ShapeFit}
\label{sec:SFresults}
To systematically quantify the impact of different fiducial cosmology choices, we perform SF analyses using five distinct cosmologies at the level of compressed parameters. These cosmologies define the \textit{grid} and \textit{template} cosmologies used in our analysis and correspond to the same set of cosmologies explored in the FM approach: \texttt{baseline, low-$\Omega_\mathrm{m}$, thawing-DE, high-$N_\mathrm{eff}$, DESI BAO}. In addition, we further include a sixth cosmology \texttt{low-$\sigma_8$} which only impacts the construction of the template. 
Unless otherwise stated, both the \textit{grid} and \textit{template} are consistently adjusted to reflect the different choices in fiducial cosmology.

We quantify the shifts using Eq.~\eqref{eq:shift} where $\Delta x$ represents the difference between the respective secondary cosmologies tested and the DESI \texttt{baseline} choice. The set of compressed parameters considered includes $\{ \alpha_\parallel, \alpha_\perp, f\sigma_{s8},\, (m+n)\}$. 
As discussed in the following, the reparametrisation of the variable $(m+n)$ is motivated by the strong correlation between $m$ and 
$n$, as shown in \citep{Brieden_2021JCAP}. To properly account for the resulting degeneracy and uncertainty, we fix one of the two parameters -- specifically, $n$ -- and adopt this reparametrised form in our analysis. Following the same approach as in the FM analysis, we calculate shifts in terms of the precision of $V_{25}$, as defined in Eq.~\eqref{eq:shiftV25}. A shift is considered statistically significant if it exceeds $2\sigma_\mathrm{V25}$.

\begin{figure}[htb]
    \centering
    \includegraphics[width=1.0\linewidth]{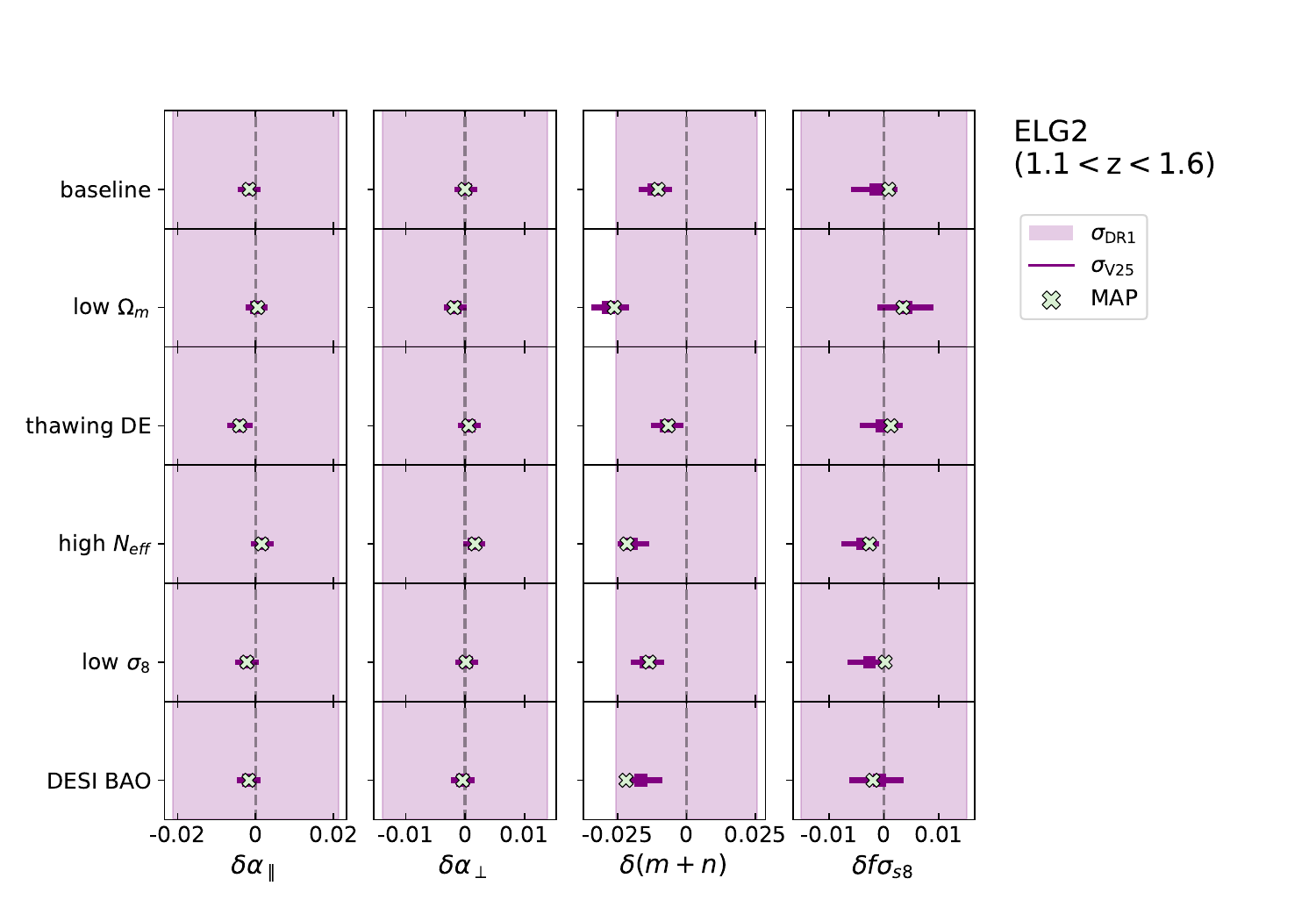}
    \caption{Same as Figure~\ref{fig:FM_ELG} but for the SF analysis, showing the 1D marginalised posterior distributions for the ELG2 sample across the full set of fiducial cosmologies studied: \texttt{baseline, low-$\Omega_\mathrm{m}$, thawing-DE, high-$N_\mathrm{eff}$, low-$\sigma_8$, DESI BAO}. Purple squares indicate the mean of the marginalised posteriors, with purple horizontal lines representing the corresponding $1\sigma_\mathrm{V25}$ error bars derived from fits to the mean of 25 Abacus-2 DR1 complete mock realisations. The purple shaded region represents the expected realistic DR1 statistical uncertainty, centred on a zero shift relative to the true cosmological parameters. Green crosses indicate the MAP values. The largest deviations from the \texttt{baseline} are observed in the $(m+n)$ parameter 
    notably in the \texttt{low-$\Omega_\mathrm{m}$} and \texttt{high-$N_\mathrm{eff}$} cases. }
    \label{fig:meanvsmap}
\end{figure}

Figure~\ref{fig:sf_AS_heatmap} illustrates the deviations in compressed parameters relative to the analysis performed with the \texttt{baseline} fiducial cosmology. All shifts remain below $2\sigma_\mathrm{V_{25}}$ for all tracers and parameters, which are consistent with statistical fluctuations associated with the variance of the 25 complete Abacus-2 DR1 simulations. We therefore conclude that no statistically significant systematic shifts are present.  Additionally, we observe a trend similar to that seen in the FM analysis, where the size of the shifts increases with redshift. 
This behavior reflects the growing sensitivity of clustering measurements to the assumed fiducial \textit{grid} cosmology at higher redshifts, which leads to larger distortions in the inferred distances. Among all cases studied, the QSO sample exhibits the most significant shift, with the \texttt{low-$\Omega_\mathrm{m}$} case reaching $1.73\sigma_\mathrm{V25}$ for $\Delta (m+n)$. For all tracers, the largest shifts occur in the reparametrised SF parameter $\Delta (m+n)$, with the most significant deviations occurring in the \texttt{low-$\Omega_\mathrm{m}$} and \texttt{high-$N_\mathrm{eff}$} cases. 
Certainly, we expect the shape parameter $m$ to be the most affected by variations in the transfer function (mainly driven by $\Omega_m h^2$ in $\Lambda$CDM) as we change the \textit{template} fiducial cosmology. In the SF formalism, the variation of the transfer function is parametrised solely by $m$, which effectively changes the slope of the power spectrum around a large-scale pivot scale, $k_p$, taking the \textit{template} fiducial cosmology as a reference. In practice, a change in the parameters $\Omega_m$ or $N_{\rm eff}$ results in a more complex modification of the power spectrum transfer function than a simple change in slope; hence, a certain systematic error is introduced. However, this systematic may only be relevant when the statistical errors in the dataset are very small (as is the case for the $V_{25}$ volume), and when the true and fiducial cosmologies have substantially different transfer functions. In a more realistic scenario, most of the \textit{template} cosmologies capable of inducing such a systematic error in $m$ are ruled out by external datasets. Thus, we consider this systematic error as merely indicative of the maximum possible systematic error in our measured $m$, which is, in practice, much smaller. 
As a robustness test, in addition to MCMC chains, we compute the MAP values in a similar way to the FM approach. Figure~\ref{fig:meanvsmap} shows the comparison between the mean of the marginalised posteriors and the MAP values for the ELG2 tracer (similar results are obtained for the other tracers). The error bars are expressed in terms of $\sigma_{\mathrm{V25}}$ and the shaded regions represent the DESI DR1 errors. 
The close alignment between posterior means and MAP estimates across all compressed parameters provides a consistency check on the inferred shifts and reinforces the conclusion that these are consistent with statistical fluctuations rather than systematic biases introduced by the choice of fiducial cosmology.

\begin{figure}[htb]
    \centering
    \includegraphics[width=0.60\linewidth]{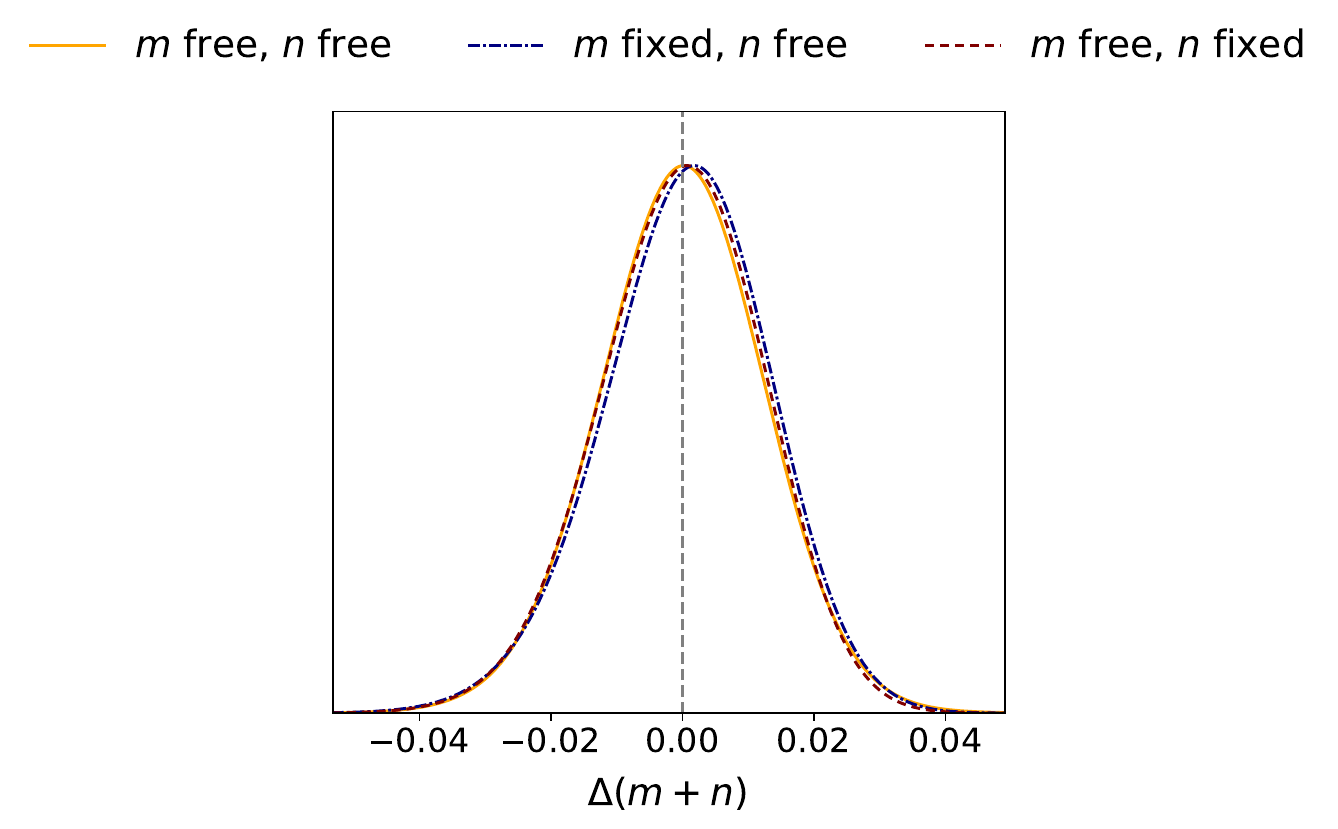}
    \includegraphics[width=0.76\linewidth]{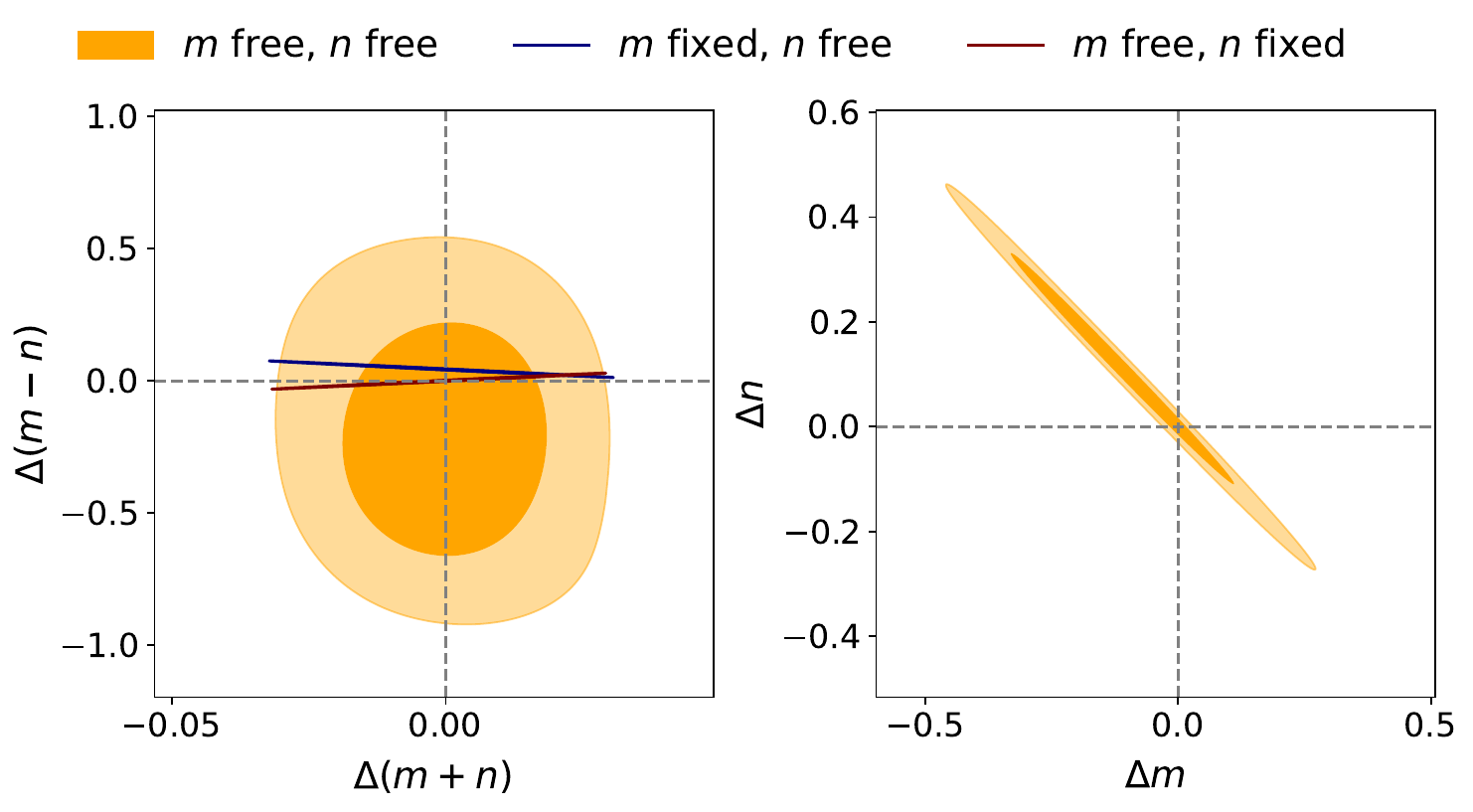}

    \caption{The posteriors for the LRG1 sample for the \texttt{high-$N_\mathrm{eff}$} cosmology. The right bottom panel presents the 2D posterior for the original $m$ and $n$ parametrisation, illustrating the strong degeneracy between the two parameters. The bottom left panel shows the posterior of the reparametrised variables $(m+n)$ and $(m-n)$. The top panel shows the 1D posterior of $\Delta(m+n)$ for the three tested scenarios, yielding consistent results across all cases. This consistency highlights the robustness of the $(m+n)$ parametrisation and justifies fixing either $m$ or $n$, and reinterpreting that variable as $m+n$, to optimize the parameter space and simplify the posterior sampling.}
    \label{fig:triangular_m_n}
\end{figure}

We now turn our attention to the reparametrisation of the variable $(m+n)$. As mentioned earlier, this choice is motivated by the strong degeneracy between $m$ and $n$. In light of this degeneracy, three scenarios are tested: {\it i)} varying both $m$ and $n$ (orange), {\it ii)} varying $m$ while fixing $n$ (burgundy), and {\it iii)} varying $n$ while fixing $m$ (navy blue); where for {\it ii)} and {\it iii)} the varied shape variable is directly reinterpreted as $m+n$. Figure~\ref{fig:triangular_m_n} shows the marginalised posteriors for the LRG1 sample in the case of the \texttt{high-$N_\mathrm{eff}$} cosmology. The bottom right panel presents the 2D posterior in the original $(m,n)$ space, where the shape parameters are strongly degenerate. In contrast, the bottom left panel shows the posteriors of the reparametrised variables for these three cases. We see that the posterior in the $m+n$ axis is well constrained and shows an excellent agreement for the three cases (see also top panel); while for case {\it i)} the posterior for the $m-n$ variable shows very loose constraints. This justifies that the information in that variable is in practice, not useful, and therefore the {\it ii)} or {\it iii)} approaches, in which this variable is ignored, can be safely adopted. The top panel displays the 1D posterior distribution of $\Delta (m+n)$ for the three cases, demonstrating that fixing either $m$ or $n$ yields perfectly consistent results. This confirms that the $(m+n)$ reparametrisation effectively captures the relevant information while mitigating the impact of the $m$ and $n$ degeneracy.

In order to report the systematic contributions associated to the SF approach, we rescale the observed shifts in terms of the expected DESI DR1 uncertainties. Table~\ref{tab:SF_max_shifts} summarises the maximum shift observed for each parameter, expressed in units of $\sigma_\mathrm{DR1}$ as defined in Eq.~\eqref{eq:DR1-error}. For most parameters, namely ${\alpha_{\parallel}, \alpha_\perp, f\sigma_{s8}}$, the shifts remain below $0.15\sigma_\mathrm{DR1}$. Slightly larger deviations are observed for the reparametrised parameter $(m+n)$, but all remain below $0.45\sigma_\mathrm{DR1}$ across all tracers. 
These shifts are within DR1's statistical uncertainty, indicating that the impact of fiducial cosmology variations is minimal.

\begin{table}[htb]
    \centering
    \begin{tabular}{|lcccc|}
        \hline
        \multicolumn{5}{|c|}{$N_{\sigma_\mathrm{DR1}}$}\\
        \hline
        \hline
        Tracer  & $\Delta \alpha_{\parallel} $ & $\Delta \alpha_{\perp} $ & $\Delta (m + n )$ & $\Delta f \sigma_{s8}$  \\
        \hline
        BGS   & 0.15  &0.06     & 0.21  &0.08 \\ 
        LRG1  & 0.08  &0.05     &0.23   &0.02 \\
        LRG2  & 0.13  &0.07     &0.18   &0.13 \\
        LRG3  & 0.15  &0.07     &0.45   &0.08 \\
        ELG2  & 0.08  &0.07     &0.31   &0.12 \\
        QSO   & 0.05  &0.10      &0.27   &0.07 \\
        \hline

    \end{tabular}
    \caption{Maximum shifts in compressed parameter for each tracer, expressed in units of the DESI DR1 statistical uncertainties, $\sigma_\mathrm{DR1}$ (see Eq.~\eqref{eq:DR1-error}). The shifts on $\alpha_{\parallel}, \alpha_\perp$ and $f\sigma_{s8}$ are below $0.15\sigma_\mathrm{DR1}$. The parameter $(m+n)$ shifts are slightly larger but below $0.45\sigma_\mathrm{DR1}$ for all tracers.
    These shifts all fall within the statistical uncertainty of DR1, indicating that variations in fiducial cosmology have a small systematic impact.}
    \label{tab:SF_max_shifts}
\end{table}

\begin{figure}[!htb]
\includegraphics[width=0.86\linewidth]{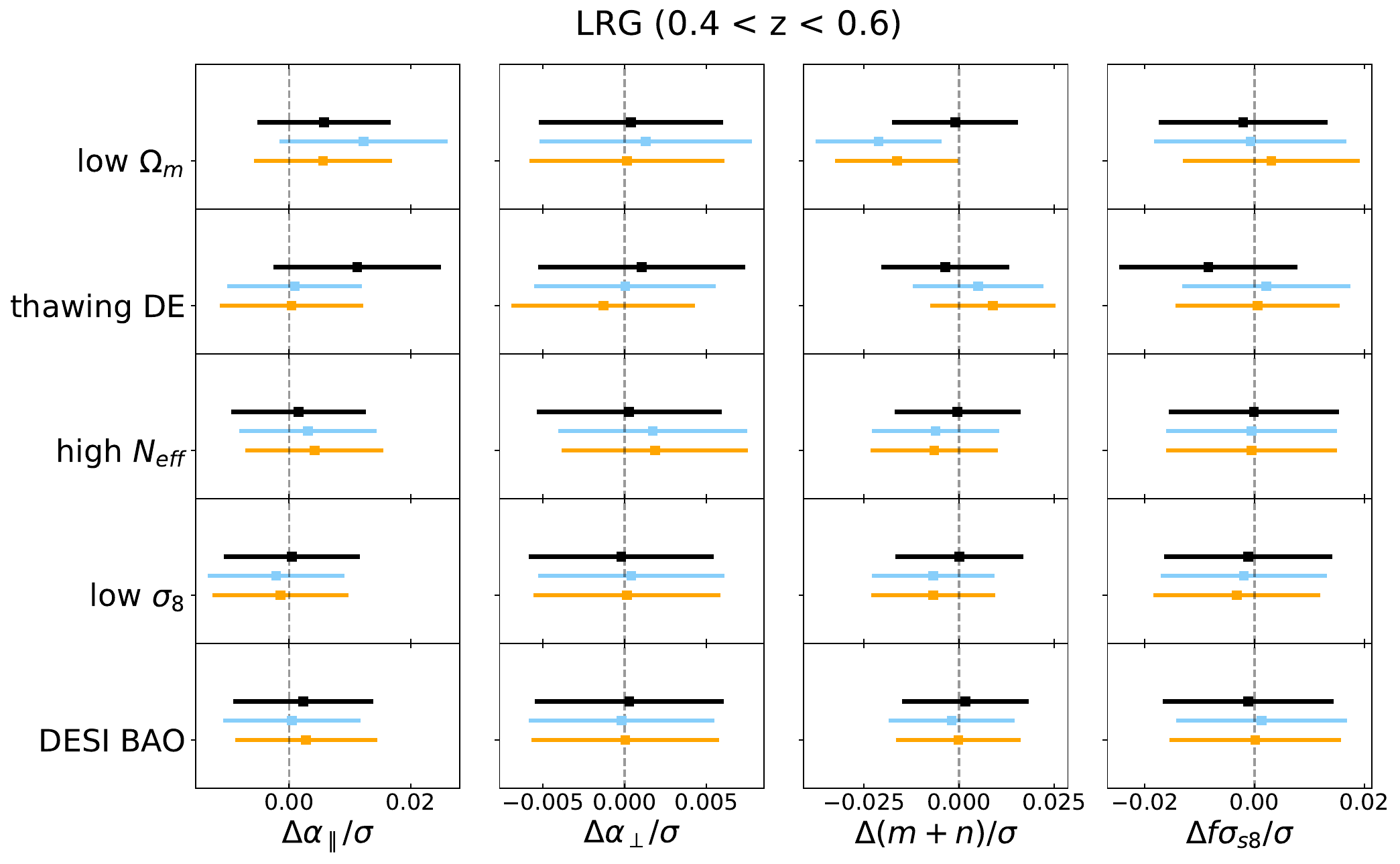}
\includegraphics[width=0.85\linewidth]{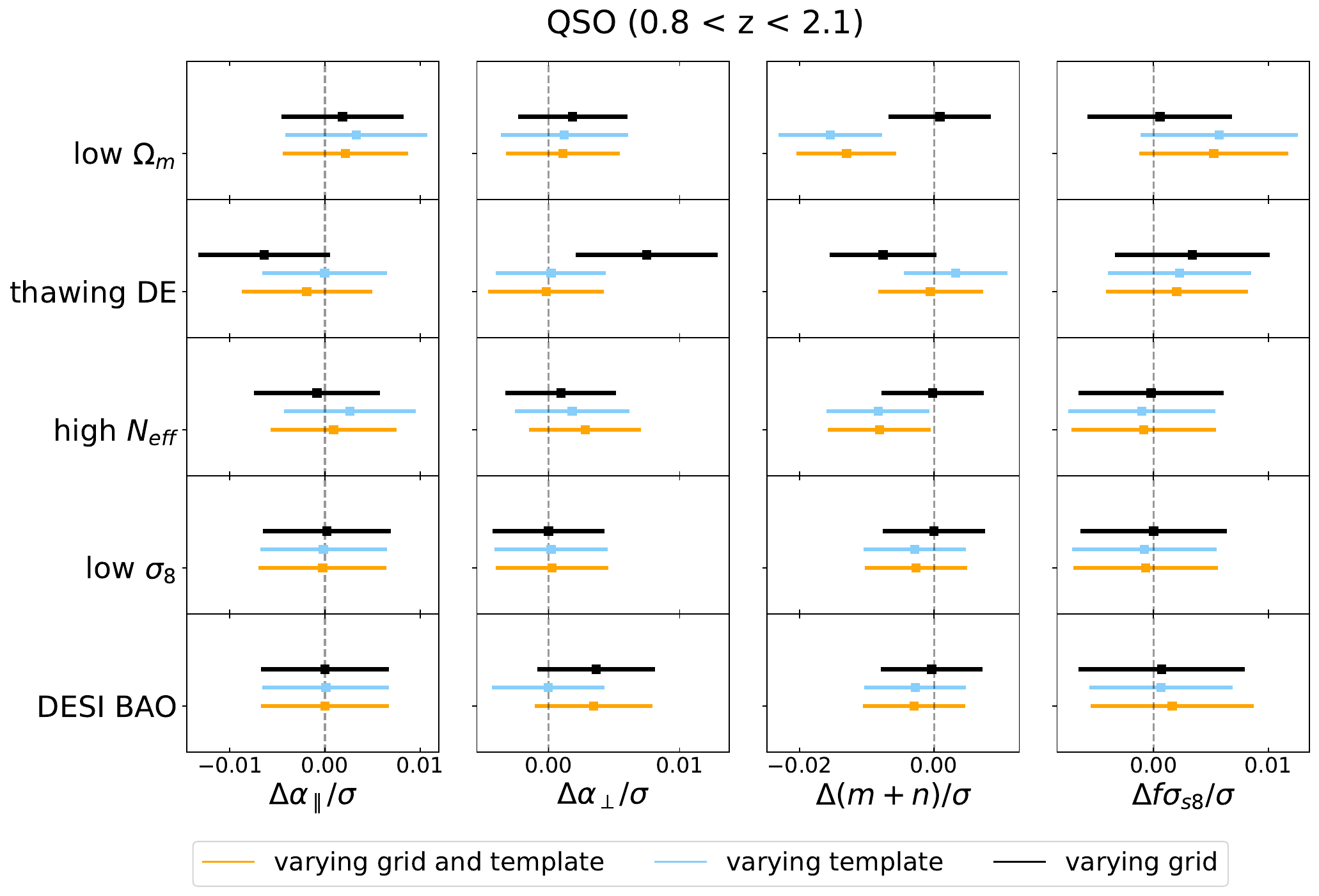}
    \caption{
    1D marginalized posterior distributions for the LRG tracer in the lowest redshift bin and the QSO tracer in the highest redshift bin, as labelled. The results are shown for three scenarios: (i) both the \textit{grid} and \textit{template} cosmologies varied consistently, (ii) the \textit{grid} cosmology varied while keeping the \textit{template} fixed, and (iii) the \textit{template} cosmology varied with a fixed \textit{grid}. For both tracers, the most significant shifts occur in the \texttt{low-$\Omega_m$} and \texttt{high-$N_{\rm eff}$} cosmologies, where the dominant bias is introduced by the mismatch in the \textit{template} cosmology. In contrast, for the \texttt{thawing-DE} and \texttt{DESI BAO} cosmologies, the main contribution to the shifts arises from the \textit{grid} cosmology, affecting the scaling parameters. In the case of \texttt{thawing-DE} cosmology, this also affects $f\sigma_{s8}$ due to its strong correlation with the Alcock-Paczynski effect.}
    \label{fig:grid_vs_template1}
\end{figure}

In addition to jointly varying the \textit{grid} and \textit{template} cosmologies, we also study the impact of changing the \textit{grid} and \textit{template} cosmologies separately by exploring two additional cases:
\begin{enumerate}
\item Changing the \textit{grid} cosmology while fixing the template cosmology to the \texttt{baseline} choice, \item Changing the \textit{template} cosmology while fixing the \textit{grid} cosmology to the \texttt{baseline} choice. 
\end{enumerate}
While the comparison of the two cases to the jointly varied case was performed for all tracers and redshift bins, Figure~\ref{fig:grid_vs_template1} illustrates the results for LRG1 and QSO, as a representation of the general trends. The figure shows the
shifts for each cosmology in the three scenarios. For both tracers, the largest shifts are observed in the \texttt{low-$\Omega_\mathrm{m}$} and \texttt{high-$N_\mathrm{eff}$} cosmologies, where the dominant bias originates from the \textit{template} cosmology. This primarily affects the $(m+n)$ parameter and, to a lesser extent, the scaling parameters, $\{\alpha_\parallel,\,\alpha_\perp\}$ via the ratio $r_\mathrm{d}^{\mathrm{temp.}}/r_\mathrm{d}^{\texttt{baseline}}$, both sensitive to early-time physics, equality and recombination epochs, respectively. In contrast, for the \texttt{thawing-DE} and \texttt{DESI BAO} cosmology, the main contribution comes from the \textit{grid} cosmology, impacting mainly the scaling parameters and $f\sigma_{s8}$. 
Similar trends in the scaling parameters were also observed in the companion BAO analysis~\citep{KP4s9-Perez-Fernandez}, where the \texttt{high-$N_\mathrm{eff}$} cosmology showed shifts related to the \textit{template}, and the \texttt{thawing-DE} case exhibited large dispersion due to the underlying \textit{grid} cosmology.
In the SF (as well as in the FM) approach, we approximate that the whole change in expansion history {\it within each $z$-bin} is described by just the two dilation parameters, $\alpha_\parallel,\,\alpha_\perp$. However, we know that this is only true in the limit where the $z$-bin is sufficiently narrow. For sufficiently wide bins, extreme cosmologies (like the \texttt{thawing-DE} or \texttt{DESI BAO} case) can cause changes in the distance-dilations within the $z$-bin that are not well captured by these $\alpha_\parallel,\,\alpha_\perp$ parameterisation. In the case of \texttt{thawing-DE} cosmology, this also affects the $f\sigma_{s8}$ parameter, due to its strong correlation with the Alcock-Paczynski effect, and with the systematic effect in the dilation of the reference scale for the smoothing. 
For \texttt{low-$\sigma_8$}, the shifts are smaller—below $0.5\sigma_{V25}$—and primarily affect $(m+n)$ and $f\sigma_{s8}$, related with variations in the \textit{template} cosmology.

\clearpage

\section{Conclusion}
\label{sec:Conclusion}
In this study, we assessed the systematic error budget associated with the choice of fiducial cosmology in the DESI full-shape analysis. To this end, we utilised the 25 realisations of the `complete' Abacus-2 DR1 mock catalogues, which are specifically tailored to the survey realism of the different DESI tracers (BGS, LRG, ELG, and QSO). These mocks, based on the Planck 2018 \lcdm{} best-fit cosmology, allowed us to quantify potential biases in cosmological parameters arising from the assumption of fiducial cosmologies that differ from the true underlying cosmology of the mocks. Our analysis considered the primary DESI \texttt{baseline} cosmology, which aligns with the Planck 2018 \lcdm{} best-fit model, along with four secondary Abacus cosmologies and the DESI DR1 BAO best-fit $w_0w_a$ cosmology. We evaluated the impact of varying the fiducial cosmology on two different full-shape modelling approaches: the full-modelling (FM) method and the ShapeFit (SF) method. While FM directly infers cosmological parameters within a given model, SF compresses information into a set of derived parameters based on a template-based framework. In both cases, the assumed fiducial cosmology affects the analysis through the distance-redshift relation (\textit{grid} cosmology), and in the case of SF, it additionally enters through the construction of the template (\textit{template} cosmology). To assess these effects, we examined four secondary \textit{grid} cosmologies—\texttt{low-$\Omega_\mathrm{m}$}, \texttt{thawing-DE}, \texttt{high-$N_\mathrm{eff}$}, and \texttt{DESI BAO}—and, in the case of the SF \textit{template} cosmology, an additional fifth fiducial cosmology, \texttt{low-$\sigma_8$}. We quantified systematic shifts in cosmological parameters with regards to the DESI \texttt{baseline} choice and reported them in units of the uncertainty from the  25 Abacus mock realisations ($\sigma_\mathrm{V25}$). A systematic shift was considered statistically significant if it exceeded the threshold of $> 2 \sigma_\mathrm{V25}$. Finally, we reported the overall systematic error budget associated with the fiducial cosmology assumption in terms of the realistic DESI DR1 statistical uncertainties.

For the FM approach, we first investigated \lcdm{}-like fiducial cosmologies (\texttt{low-$\Omega_\mathrm{m}$} and \texttt{high-$N_\mathrm{eff}$})  and found that systematic shifts were small (below 1$\sigma_{V25}$) across all tracers, demonstrating that the FM analysis is robust to moderate variations in the fiducial cosmology.  Within the joint analysis of all tracers, the shifts remained well within the statistical uncertainties of DESI DR1, with the largest detected shift reaching $0.22\sigma_\mathrm{DR1}$. In a second step, we tested two beyond-\lcdm{} cosmology (\texttt{thawing DE} and \texttt{DESI BAO}) with additional degrees of freedom in $w_0$ and $w_a$. As a stress test, we first performed FM inference assuming these strongly non-\lcdm{} cosmologies while still sampling a standard \lcdm{} model. This revealed significantly biased cosmological parameters, demonstrating the limitations of a restricted model in capturing the effects of an imposed beyond \lcdm{} \textit{grid} cosmology. Subsequently, we performed an analysis of these extended \textit{grid} cosmologies in a fully consistent beyond \lcdm{} framework. Due to parameter degeneracies and projection effects, the DESI full-shape data alone lack the statistical power to robustly constrain this extended parameter space. To mitigate these effects, we incorporated SNe Ia mock data to obtain a realistic and stable estimate of the systematic shifts.  In this combined analysis, all shifts in key cosmological parameters remain small, below $0.12\sigma_\mathrm{DR1+SN}$. We conclude that the systematic impact of the fiducial cosmology is negligible for both \lcdm{} and beyond-\lcdm{} models when full-shape data are combined with complementary probes. This robustness is expected to improve further with the inclusion of DESI BAO or CMB data. Given that the official DESI DR1 FM results~\citep{DESI2024.VII.KP7B} for $w_0w_a$CDM are presented in combination with BAO, CMB, and SNe Ia data, we conclude that the systematic contribution from the choice of fiducial cosmology is negligible compared to the overall statistical precision of DR1.

For the SF approach, we studied the effect of simultaneously and consistently changing the grid and template cosmology across the five secondary cosmologies, relative to the DESI \texttt{baseline} choice. We find that, for all individual tracers, the systematic shifts remain below the $2\sigma_\mathrm{V25}$ threshold, with the largest deviation detected in the reparameterised SF combination $\Delta(m+n)$, reaching $1.73 \sigma_\mathrm{V25}$. The most pronounced shifts are associated with the \texttt{low-$\Omega_\mathrm{m}$} and \texttt{high-$N_\mathrm{eff}$} cosmologies, where the fixed value of the shape parameter $n$ does not match its true underlying value in the mocks. 
Compared with the statistical error of DR1, the shifts observed on most of the parameters, $\{\alpha_{\parallel}, \alpha_\perp, f\sigma_{s8}\}$, remain below $0.15\sigma_\mathrm{DR1}$ and below $0.45\sigma_\mathrm{DR1}$, for $(m+n)$ for all tracers. All the shifts remain within the statistical uncertainty of DR1. We also investigated the separate contribution of the \textit{template} and the \textit{grid} cosmologies, observing that the most significant shifts occur in the \texttt{low-\(\Omega_\mathrm{m}\)} and \texttt{high-\(N_\mathrm{eff}\)} cosmologies. Here, the largest shifts originate from the \textit{template} cosmology, primarily impacting the parameter $(m+n)$ and, to a lesser extent, the scaling parameters via the sound-horizon ratio, both influenced by early Universe evolution. In contrast, in the \texttt{thawing DE} and \texttt{DESI BAO} scenarios, the \textit{grid} cosmology contributes mainly, affecting the scaling parameters through the late-time expansion history, such as the distance-redshift relation, and \(f\sigma_{s8}\).

Our results are presented in Section 5 of~\citep{DESI2024.V.KP5}, where we conclude that, at the precision level of DESI DR1, the full-shape analysis—both in the FM and SF approaches—remains robust against variations in fiducial cosmology. Given that the impact of changing the fiducial cosmology is minor compared to the statistical precision of DR1, no additional contribution from this source was included in the total systematic error budget. While the deviation observed in the \texttt{low-$\Omega_m$} and \texttt{high-$N_\mathrm{eff}$} cases reaches up to $\sim 0.45\sigma_\mathrm{DR1}$ in the $(m+n)$ combination, we emphasise that this reflects the most pessimistic scenario, and that these cosmologies are not representative of viable cosmologies within the DESI parameter space. Therefore, we do not propagate this shift as a systematic uncertainty into the final $(m+n)$ result. However, as the statistical precision of DESI improves with future data releases, it will become increasingly important to revisit the assessment of systematic uncertainties due to the choice of fiducial cosmology. In particular, if evidence for evolving dark energy persists in the near future, it will be crucial to expand the set of beyond-\lcdm{} cosmologies explored in this context. Although no mitigation strategy was required for DR1 given the subdominant nature of this systematic, a natural approach in the future would be to propagate any systematic shifts into the total error budget, as done in previous RSD analyses (see e.g.~\cite{deMattiaeBOSS2020}). Alternatively, one could update the fiducial grid cosmology if the scaling parameters $\alpha_\parallel,\,\alpha_\perp$ are found to deviate substantially from unity. Extensions beyond the standard linear rescaling (e.g. redshift-dependent dilation parameters) are in principle possible but would add complexity without clear benefit at the current precision level. While our current analysis relies on mocks generated from a single underlying true cosmology, with the fiducial cosmology varied systematically within the analysis pipeline, it will also be valuable to investigate the complementary scenario of analysing mock catalogues generated from different underlying cosmologies with a fixed fiducial cosmology. Despite these considerations, the current analysis demonstrates that the fiducial cosmology choice has a negligible impact on the cosmological inference, reinforcing the robustness of the full-shape analysis with DESI DR1 data.

\section*{Data Availability}
Data from the plots in \href{https://zenodo.org/records/17087040}{this paper}\nocite{Zenodo} will be available on Zenodo as part of DESI’s Data Management
Plan.

\acknowledgments
RG  acknowledges support from the Swiss National Science Foundation (SNF) ``Cosmology
with 3D Maps of the Universe" research grant, 200020\_175751 and 200020\_207379. MV and SR acknowledge PAPIIT-IN108321, PAPIIT-IA103421, PAPIIT-IN116024 and PAPIIT-IN115424. SR acknowledges support from the Universidad Nacional Autónoma de México Postdoctoral Program (POSDOC). HGM acknowledges support through the programmes Ramón y Cajal (RYC-2021-034104) and Consolidación Investigadora (CNS2023-144605) of the Spanish Ministry of Science and Innovation. SR and MV acknowledge Sebastien Fromenteau for the thorough discussions about the methodology.\\
\indent This material is based upon work supported
by the U.S. Department of Energy (DOE), Office of Science, Office of High-Energy Physics,
under Contract No. DE–AC02–05CH11231, and by the National Energy Research Scientific
Computing Center, a DOE Office of Science User Facility under the same contract. Additional
support for DESI was provided by the U.S. National Science Foundation (NSF), Division of
Astronomical Sciences under Contract No. AST-0950945 to the NSF’s National OpticalInfrared Astronomy Research Laboratory; the Science and Technology Facilities Council of
the United Kingdom; the Gordon and Betty Moore Foundation; the Heising-Simons Foundation; the French Alternative Energies and Atomic Energy Commission (CEA); the National Council of Humanities, Science and Technology of Mexico (CONAHCYT); the Ministry of Science and Innovation of Spain (MICINN), and by the DESI Member Institutions:
\url{https://www.desi.lbl.gov/collaborating-institutions}. 
The DESI Legacy Imaging Surveys consist of three individual and complementary projects: the Dark Energy
Camera Legacy Survey (DECaLS), the Beijing-Arizona
Sky Survey (BASS), and the Mayall z-band Legacy Survey (MzLS). DECaLS, BASS and MzLS together include data obtained, respectively, at the Blanco telescope, Cerro Tololo Inter-American Observatory, NSF
NOIRLab; the Bok telescope, Steward Observatory, University of Arizona; and the Mayall telescope, Kitt Peak
National Observatory, NOIRLab. NOIRLab is operated
by the Association of Universities for Research in Astronomy (AURA) under a cooperative agreement with
the National Science Foundation. Pipeline processing
and analyses of the data were supported by NOIRLab
and the Lawrence Berkeley National Laboratory. Legacy
Surveys also uses data products from the Near-Earth Object Wide-field Infrared Survey Explorer (NEOWISE),
a project of the Jet Propulsion Laboratory/California Institute of Technology, funded by the National Aeronautics and Space Administration. Legacy Surveys was
supported by: the Director, Office of Science, Office of
High Energy Physics of the U.S. Department of Energy; the National Energy Research Scientific Computing Center, a DOE Office of Science User Facility; the
U.S. National Science Foundation, Division of Astronomical Sciences; the National Astronomical Observatories of China, the Chinese Academy of Sciences and the
Chinese National Natural Science Foundation. LBNL
is managed by the Regents of the University of California under contract to the U.S. Department of Energy. The complete acknowledgments can be found at~\url{https://www.legacysurvey.org/}.
Any opinions, findings, and
conclusions or recommendations expressed in this material are those of the author(s) and
do not necessarily reflect the views of the U. S. National Science Foundation, the U. S.
Department of Energy, or any of the listed funding agencies.\\
\indent The authors are honored to be permitted to conduct scientific research on Iolkam Du’ag
(Kitt Peak), a mountain with particular significance to the Tohono O’odham Nation.

\appendix
\section{Appendix: Table DR1-like errors} \label{app:appendix}
To complement the discussion in Section~\ref{sec:DR1error}, we report here the DR1-like statistical uncertainties, $\sigma_\mathrm{DR1}$, for the cosmological and compressed parameters used in this work. These uncertainties are derived following the procedure outlined in the main text, using the FFA EZmock covariance matrices including the effects of fiber assignment and the $\theta$-cut. For cosmological parameters in the extended  $w_0w_a$CDM, we additionally report the statistical uncertainties obtained when combining DESI DR1 FS data with a SNe Ia mock dataset (see Section~\ref{sec:externaldata}), denoted $\sigma_\mathrm{DR1+SN}$.

\begin{table}[htb]
\centering
\begin{tabular}{|lccccccccc|}
    \hline
    \multicolumn{10}{|c|}{$\sigma_\mathrm{DR1}$}\\
    \hline
    Tracer  & $h$ & $ \omega_{\rm cdm}$ & $ \ln (10^{10} A_s)$& $ w_0$ & $ w_a$ & $\alpha_\parallel$ & $\alpha_\perp$ & (m+n) & $f \sigma_8$\\
    \hline
    BGS     & 0.0443  & 0.0199  &  0.2604  & -- & -- & 0.0700 & 0.0401 & 0.0966 &  0.0807\\
    LRG1  & 0.0274  & 0.0144 & 0.2356 & -- &  -- &     0.0602 & 0.0290& 0.0730& 0.0659\\
    LRG2  &0.0199   & 0.0124  & 0.2160 & --& -- &      0.0469 & 0.0218& 0.0601& 0.0509\\
    LRG3  & 0.0175  & 0.0113  &  0.2056 & -- & -- & 0.0381    & 0.0198& 0.0504& 0.0461\\
    ELG2  & 0.0307  & 0.0108  & 0.1927 & -- & -- & 0.0425     & 0.0279& 0.0513& 0.0302\\
    QSO  & 0.0292  & 0.0096  & 0.1883 & -- & -- & 0.0412      & 0.0289& 0.0474& 0.0383\\
    combined & 0.0104& 0.0057 & 0.0895 & -- & -- & -- & -- &  --& -- \\
    \hline
    \hline
    \multicolumn{10}{|c|}{$\sigma_\mathrm{DR1+SN}$}\\
    \hline
    combined & 0.0116&  0.0056&  0.1641&  0.0817 &  0.4665& --  & --& --&-- \\
    \hline
\end{tabular}
\caption{DR1-like ($\sigma_\mathrm{DR1}$) and combined DR1+SN ($\sigma_\mathrm{DR1+SN}$) $1\sigma$ statistical uncertainties on cosmological and compressed parameters used in this work.}
    \label{tab:AppendixA}
\end{table}

\clearpage

\section{Author Affiliations}
\label{sec:affiliations}

\noindent \hangindent=.5cm $^{1}${Institute of Physics, Laboratory of Astrophysics, \'{E}cole Polytechnique F\'{e}d\'{e}rale de Lausanne (EPFL), Observatoire de Sauverny, Chemin Pegasi 51, CH-1290 Versoix, Switzerland}

\noindent \hangindent=.5cm $^{2}${Instituto de F\'{\i}sica, Universidad Nacional Aut\'{o}noma de M\'{e}xico,  Circuito de la Investigaci\'{o}n Cient\'{\i}fica, Ciudad Universitaria, Cd. de M\'{e}xico  C.~P.~04510,  M\'{e}xico}

\noindent \hangindent=.5cm $^{3}${Instituto de Ciencias F\'{\i}sicas, Universidad Aut\'onoma de M\'exico, Cuernavaca, Morelos, 62210, (M\'exico)}

\noindent \hangindent=.5cm $^{4}${Institut de Ci\`encies del Cosmos (ICCUB), Universitat de Barcelona (UB), c. Mart\'i i Franqu\`es, 1, 08028 Barcelona, Spain.}

\noindent \hangindent=.5cm $^{5}${Institute of Cosmology and Gravitation, University of Portsmouth, Dennis Sciama Building, Portsmouth, PO1 3FX, UK}

\noindent \hangindent=.5cm $^{6}${Departament de F\'{\i}sica Qu\`{a}ntica i Astrof\'{\i}sica, Universitat de Barcelona, Mart\'{\i} i Franqu\`{e}s 1, E08028 Barcelona, Spain}

\noindent \hangindent=.5cm $^{7}${Institut d'Estudis Espacials de Catalunya (IEEC), c/ Esteve Terradas 1, Edifici RDIT, Campus PMT-UPC, 08860 Castelldefels, Spain}

\noindent \hangindent=.5cm $^{8}${Sorbonne Universit\'{e}, CNRS/IN2P3, Laboratoire de Physique Nucl\'{e}aire et de Hautes Energies (LPNHE), FR-75005 Paris, France}

\noindent \hangindent=.5cm $^{9}${IRFU, CEA, Universit\'{e} Paris-Saclay, F-91191 Gif-sur-Yvette, France}

\noindent \hangindent=.5cm $^{10}${Institute for Astronomy, University of Edinburgh, Royal Observatory, Blackford Hill, Edinburgh EH9 3HJ, UK}

\noindent \hangindent=.5cm $^{11}${Max Planck Institute for Extraterrestrial Physics, Gießenbachstraße 1, 85748 Garching,Germany}

\noindent \hangindent=.5cm $^{12}${Lawrence Berkeley National Laboratory, 1 Cyclotron Road, Berkeley, CA 94720, USA}

\noindent \hangindent=.5cm $^{13}${Department of Physics, Boston University, 590 Commonwealth Avenue, Boston, MA 02215 USA}

\noindent \hangindent=.5cm $^{14}${Dipartimento di Fisica ``Aldo Pontremoli'', Universit\`a degli Studi di Milano, Via Celoria 16, I-20133 Milano, Italy}

\noindent \hangindent=.5cm $^{15}${INAF-Osservatorio Astronomico di Brera, Via Brera 28, 20122 Milano, Italy}

\noindent \hangindent=.5cm $^{16}${Department of Physics \& Astronomy, University College London, Gower Street, London, WC1E 6BT, UK}

\noindent \hangindent=.5cm $^{17}${Institute of Space Sciences, ICE-CSIC, Campus UAB, Carrer de Can Magrans s/n, 08913 Bellaterra, Barcelona, Spain}

\noindent \hangindent=.5cm $^{18}${NSF NOIRLab, 950 N. Cherry Ave., Tucson, AZ 85719, USA}

\noindent \hangindent=.5cm $^{19}${Institut de F\'{i}sica d’Altes Energies (IFAE), The Barcelona Institute of Science and Technology, Edifici Cn, Campus UAB, 08193, Bellaterra (Barcelona), Spain}

\noindent \hangindent=.5cm $^{20}${Departamento de F\'isica, Universidad de los Andes, Cra. 1 No. 18A-10, Edificio Ip, CP 111711, Bogot\'a, Colombia}

\noindent \hangindent=.5cm $^{21}${Observatorio Astron\'omico, Universidad de los Andes, Cra. 1 No. 18A-10, Edificio H, CP 111711 Bogot\'a, Colombia}

\noindent \hangindent=.5cm $^{22}${University of Virginia, Department of Astronomy, Charlottesville, VA 22904, USA}

\noindent \hangindent=.5cm $^{23}${Fermi National Accelerator Laboratory, PO Box 500, Batavia, IL 60510, USA}

\noindent \hangindent=.5cm $^{24}${Steward Observatory, University of Arizona, 933 N. Cherry Avenue, Tucson, AZ 85721, USA}

\noindent \hangindent=.5cm $^{25}${Institut d'Astrophysique de Paris. 98 bis boulevard Arago. 75014 Paris, France}

\noindent \hangindent=.5cm $^{26}${Center for Cosmology and AstroParticle Physics, The Ohio State University, 191 West Woodruff Avenue, Columbus, OH 43210, USA}

\noindent \hangindent=.5cm $^{27}${Department of Physics, The Ohio State University, 191 West Woodruff Avenue, Columbus, OH 43210, USA}

\noindent \hangindent=.5cm $^{28}${The Ohio State University, Columbus, 43210 OH, USA}

\noindent \hangindent=.5cm $^{29}${School of Mathematics and Physics, University of Queensland, Brisbane, QLD 4072, Australia}

\noindent \hangindent=.5cm $^{30}${Department of Physics, University of Michigan, 450 Church Street, Ann Arbor, MI 48109, USA}

\noindent \hangindent=.5cm $^{31}${University of Michigan, 500 S. State Street, Ann Arbor, MI 48109, USA}

\noindent \hangindent=.5cm $^{32}${Department of Physics, The University of Texas at Dallas, 800 W. Campbell Rd., Richardson, TX 75080, USA}

\noindent \hangindent=.5cm $^{33}${Department of Physics, Southern Methodist University, 3215 Daniel Avenue, Dallas, TX 75275, USA}

\noindent \hangindent=.5cm $^{34}${Department of Physics and Astronomy, University of California, Irvine, 92697, USA}

\noindent \hangindent=.5cm $^{35}${Center for Astrophysics $|$ Harvard \& Smithsonian, 60 Garden Street, Cambridge, MA 02138, USA}

\noindent \hangindent=.5cm $^{36}${Departament de F\'{i}sica, Serra H\'{u}nter, Universitat Aut\`{o}noma de Barcelona, 08193 Bellaterra (Barcelona), Spain}

\noindent \hangindent=.5cm $^{37}${Instituci\'{o} Catalana de Recerca i Estudis Avan\c{c}ats, Passeig de Llu\'{\i}s Companys, 23, 08010 Barcelona, Spain}

\noindent \hangindent=.5cm $^{38}${Department of Physics and Astronomy, Siena College, 515 Loudon Road, Loudonville, NY 12211, USA}

\noindent \hangindent=.5cm $^{39}${Department of Physics and Astronomy, University of Sussex, Brighton BN1 9QH, U.K}

\noindent \hangindent=.5cm $^{40}${Department of Physics and Astronomy, University of Waterloo, 200 University Ave W, Waterloo, ON N2L 3G1, Canada}

\noindent \hangindent=.5cm $^{41}${Perimeter Institute for Theoretical Physics, 31 Caroline St. North, Waterloo, ON N2L 2Y5, Canada}

\noindent \hangindent=.5cm $^{42}${Waterloo Centre for Astrophysics, University of Waterloo, 200 University Ave W, Waterloo, ON N2L 3G1, Canada}

\noindent \hangindent=.5cm $^{43}${Instituto de Astrof\'{i}sica de Andaluc\'{i}a (CSIC), Glorieta de la Astronom\'{i}a, s/n, E-18008 Granada, Spain}

\noindent \hangindent=.5cm $^{44}${Departament de F\'isica, EEBE, Universitat Polit\`ecnica de Catalunya, c/Eduard Maristany 10, 08930 Barcelona, Spain}

\noindent \hangindent=.5cm $^{45}${Department of Physics and Astronomy, Sejong University, 209 Neungdong-ro, Gwangjin-gu, Seoul 05006, Republic of Korea}

\noindent \hangindent=.5cm $^{46}${Abastumani Astrophysical Observatory, Tbilisi, GE-0179, Georgia}

\noindent \hangindent=.5cm $^{47}${Department of Physics, Kansas State University, 116 Cardwell Hall, Manhattan, KS 66506, USA}

\noindent \hangindent=.5cm $^{48}${Faculty of Natural Sciences and Medicine, Ilia State University, 0194 Tbilisi, Georgia}

\noindent \hangindent=.5cm $^{49}${CIEMAT, Avenida Complutense 40, E-28040 Madrid, Spain}

\noindent \hangindent=.5cm $^{50}${Department of Physics \& Astronomy, Ohio University, 139 University Terrace, Athens, OH 45701, USA}

\noindent \hangindent=.5cm $^{51}${Department of Astronomy, Tsinghua University, 30 Shuangqing Road, Haidian District, Beijing, China, 100190}

\noindent \hangindent=.5cm $^{52}${National Astronomical Observatories, Chinese Academy of Sciences, A20 Datun Road, Chaoyang District, Beijing, 100101, P.~R.~China}

\clearpage
\bibliographystyle{JHEP}
\bibliography{refFidCos}

\end{document}